\RequirePackage{fix-cm}
\documentclass[twocolumn,epjc3]{svjour3}  
\smartqed  
\usepackage{subcaption}
\RequirePackage{graphicx}
\RequirePackage{amsmath}
\usepackage{amssymb}
\usepackage{bigints}
\usepackage{adjustbox}
\usepackage{multirow}
\usepackage{array}
\usepackage{float}
\usepackage{cite}
\usepackage{balance}
\usepackage{rotating}

\usepackage[colorlinks=true,linkcolor=blue, citecolor=blue, urlcolor=blue]{hyperref}

\usepackage{orcidlink}
\usepackage{cite}
\usepackage{caption}
\usepackage{subcaption}
\usepackage{graphicx, amsmath}
\usepackage{float}
\usepackage{booktabs}
\usepackage{multirow}
\begin{document}

\title{Multi-Spin Perturbations, Thermodynamics, and Observational Signatures of Reissner–Nordström Black Holes in Bumblebee Gravity
}


\author{Jayasri Choudhury\orcidlink{0009-0003-2016-7596}
\thanksref{e1,addr1} 
        \and Yenshembam Priyobarta Singh\orcidlink{0009-0007-7776-9511}\thanksref{e2,addr1}
         Dhruba Jyoti Gogoi \orcidlink{}   \thanksref{e3,addr2,address3}
        Telem Ibungochouba Singh\orcidlink{0000-0002-2568-0343}\thanksref{e4,addr1}
}

\thankstext{e1}{e-mail:jayasrichou57@gmail.com }
\thankstext{e2}{e-mail:  priyoyensh@gmail.com   }
\thankstext {e3}{e-mail: moloydhruba@yahoo.in   }
\thankstext{e4}{e-mail: ibungochouba@rediffmail.com (corresponding author)}


\institute{Department of Mathematics, Manipur University, Canchipur 795003, Manipur, India \label{addr1}
          \and
                   Department of Physics, Madhabdev University, Lakhimpur, 784164, Assam, India\label{addr2}
                   \and
                   Research Center of Astrophysics and Cosmology, Khazar University, Baku, AZ1096, 41, Mehseti Street, Azerbaijan\label{address3}
}

\date{Received: date / Accepted: date}

\maketitle

\begin{abstract}
In this paper, we present a comprehensive investigation into the dynamical and thermodynamic properties of the charged Reissner-Nordstr\"{o}m (RN) black hole (BH) in the  bumblebee gravity framework, where spontaneous Lorentz symmetry breaking (LSB) occurs.  To analyze the dynamical behavior, we apply the unified Teukolsky master equation within the Newman-Penrose formalism to evaluate massless field perturbations of arbitrary spins ($s=0, 1/2, 1, 3/2, 2$). By deriving the corresponding effective potentials, we compute the quasinormal modes (QNMs) frequencies using the Pad\'{e}-improved 6th-order WKB approximation and the Asymptotic Iteration Method (AIM) and evaluate the greybody factors for all perturbing fields, demonstrating how the LSB  parameter $L$ and the BH charge $Q$ modify the spacetime's damped oscillations and wave propagation. We further assess the observational prospects of these QNMs by determining the black-hole mass ranges accessible to current and future gravitational-wave detectors, including LISA, Virgo, and LIGO. Moreover, we investigate the modified thermodynamic structure, calculating the Hawking temperature, entropy with logarithmic thermal corrections and heat capacity. Our thermodynamic analysis reveals a second-order phase transition whose critical radius is heavily governed by the background charge. These combined findings provide valuable theoretical insights into how Lorentz violation affects the physical stability, thermal evolution, and phase structure of charged BHs.

\end{abstract}

\date{}
\section{Introduction}
Einstein's general relativity formulated in 1915, stands as one of the most successful theories of modern physics explaining the precession of Mercury's perihelion to the bending of light around massive bodies. Among these, existence of BH is the most remarkable prediction of general relativity (GR), enclosed by an event horizon beyond which nothing can escape. The first exact BH solution, obtained by Schwarzschild in 1916, followed by the charged RN solution, rotating Kerr and Kerr– \\ Newman BH together forming foundation of BH physics. Now it is not only limited to mathematical study, becoming a central objects in both theoretical and observational physics studies. The observational confirmations through direct detection of gravitational waves via LIGO/Virgo collaboration (2015) \cite{B.P.Abbott2016} and  imaging of the BH by the Event Horizon Telescope (2019, 2022) \cite{K.Akiyama2019, 2K.Akiyama2019},
embedded BHs as key testing grounds for the fundamental laws of physics in the strong-field regime.
 
 Despite remarkable success, GR remains incomplete to describe dark energy, dark matter, and quantum gravity effects at the Planck scale \cite{D.Colladay1997, D.Colladay1998}, which strongly motivates the study of modified gravity theories. Among the various directions explored, Lorentz symmetry breaking (LSB) has gained considerable attention, as  Lorentz symmetry is one of the most fundamental principle for both GR and the Standard model. Even though it has been tested successfully throughout different energy scales \cite{Mattingly2005, Liberati2013}, however near the Planck scale where quantum gravitational effects become significant, several theories such as non commutative geometry \cite{Carroll2001}, loop quantum gravity \cite{A. Ashtekar2021}, and String field theory \cite{Kostelecky1989} say Lorentz symmetry may break at some levels. Standard Model Extension (SME), introduced by Colladay and Kostelecky \cite{Colladay1998} provides a systematic framework to study such violations. Bumblebee Gravity model \cite{V.A. Kostelecký2001, V.A. Kostelecky2004} is one of the useful framework within SME, first proposed by Kostelecky and Samuel, which we adopt in this work. In this framework, Lorentz symmetry is spontaneously broken  through a vector field $B_\mu$, develops a nonzero vacuum expectation value, $\langle B_\mu \rangle = \ell_\mu$, leading to the spontaneous violation of Lorentz symmetry \cite{T. Ibungochouba Singh2026, Ibungochouba Singh2025, T. Ibungochouba Singh2025 2, Ningthoujam Media2025,T. Ibungochouba Singh2024 2, S. Christina2023, Y. Onika Laxmi2023, Y. Onika Laxmi2023 2, Y. Onika Laxmi2022, Y.Priyobarta Singh2022}. The coupling of this field to the Ricci tensor introduces a dimensionless LSB parameter ($L$), which modifies the background geometry of standard GR. Constructing RN solution within this framework produces a modified charged BH whose geometry, perturbation spectrum, and thermodynamic structure are affected by LSB. The static spherically symmetric bumblebee BH solution was first obtained by Casana et al. \cite{Casana2018}, and further studies have examined its shadow, gravitational lensing, and geodesic structure etc. The charged RN solution in bumblebee gravity was derived in \cite{Liu2025}. Perturbation theory is one of the most useful approaches to study the BH dynamics. Due to the perturbation by different fields around the BHs, it undergoes damped oscillations characterized by QNMs \cite{K.D. Kokkotas1999,R.A.Konoplya2011, Y. Priyobarta Sing2024, Y. Priyobarta Sing2025}, where the parameters of the background spacetime determines the whole complex frequencies and the real part of the frequency determines the oscillation rate, while the imaginary part governs the damping of the perturbations. Beyond QNMs, related BH observables such as GFs \cite{ M.Visser1999, S. Kanzi2021} and Hawking radiation spectra \cite{Don N. Page1976} provide valuable information regarding wave propagation and energy emission in curved spacetimes, which are investigated in various bumblebee BH backgrounds for scalar, electromagnetic, and Dirac fields using techniques such as the WKB approximation \cite{S.Iyer1987,B.F.Schutz1985}, continued-fraction method \cite{E.W. Leaver1985}, Pöschl– \\ Teller approximation \cite{V. Ferrari1994}, and Frobenius expansions \cite{R.A. Konoplya2011}. However, most existing studies treat different spin fields separately. The unified treatment of perturbations for arbitrary spin fields in BH spacetimes was developed using the Newman-Penrose (NP) formalism, which reduces the scalar, electromagnetic, and gravitational perturbation equations to a single master Teukolsky equation \cite{S.A.Teukolsky1972, S.A.Teukolsky1973, S.A.Teukolsky1974, W.H.Press1973}. While this unified framework has been successfully applied to the Schwarzschild BH
\cite{Fu-Wen Shua2005}, its extension to the charged RN background in bumblebee gravity remains unexplored. In this work, we extend this unified Teukolsky master equation framework to the RN BH in bumblebee gravity, for scalar ($s=0$), electromagnetic 
($s=1$), Dirac ($s=1/2$), Rarita-Schwinger ($s=3/2$) and gravitational ($s=2$) perturbations within a single consistent formalism to study perturbation dynamics. We also calculated the greybody factor (GF) for each spin value using WKB approach.

Besides the dynamical properties, studying BH thermodynamic properties is important to know the equilibrium behaviour and stability of gravitational systems. Since Lorentz symmetry violation modifies the geometry of the RN BH, it will modify its thermodynamic behaviour as well \cite{ding1,kastor1,ding2,jawad2017c,kumar2020,priyo2023,dhruba2026,S. Kanzi2019}. To investigate the impact of the modified background geometry, we study the BH's thermal properties, deriving the Hawking temperature $T_h$, which determines how Lorentz violation alters the thermal emission of the BH relative to the standard RN case. We compute the Bekenstein-Hawking entropy including corrections introduced by the bumblebee field, showing that the LSB parameter modifies the standard area law. The heat capacity $C_Q$ is analysed, showing a second-order phase transition at a critical radius $r_c$ that depends explicitly on both $L$ and $Q$, proving that bumblebee gravity directly controls the thermodynamic stability structure of the charged BH.

In this work, extending the unified Teukolsky master equation framework to the RN BH in bumblebee gravity, we investigate arbitrary-spin ($s= 0, 1/2, 1, 3/2, 2$) perturbations within a single consistent framework. Corresponding effective potentials are derived and QNMs are computed to examine the influence of the $L$,  $Q$, and multipole number $(l)$ on the oscillation frequencies and damping rates of different spin fields. Moreover, we analyse the thermodynamic properties such as Hawking temperature, entropy, Gibbs free energy and heat capacity. Results obtained from our study provide a unified structure to analyse the impact of $L$ in both dynamical and thermodynamic properties in modified BH spacetime.   

The organisation of the paper is as follows. In section 2, we derive the effective potentials for scalar, Dirac, electromagnetic, Rarita-Schwinger and gravitational  perturbations. In section 3, we investigate the QNMs of different field perturbations using AIM method and WKB with Padé approximation for varying $L$ and $Q$. The observational prospects of QNMs in bumblebee gravity are discussed in section 4. Thermodynamic properties of BH including  entropy, Gibbs free energy, heat capacity and the GF for different spins are discussed in section 5.  Some discussions and conclusions are given in section 6.

\section{Reissner-Nordstrom black hole }
As described in \cite{Casana2018}, the line element characterizing a charged BH configuration under bumblebee gravity is given by
\begin{align} \label{eqn 1}
	ds^2= A(r) dt^2 -\dfrac{1+L}{A(r)} dr^2-r^2 (d\theta^2+\sin^2\theta d\phi^2),
\end{align}
where
\begin{align}\label{eqn 2}
A(r)=1-\dfrac{2M}{r}+ \dfrac{2(1+L)}{2+L}\dfrac{Q^2}{r^2}.
\end{align}
Here $M$ represents the mass, $Q$ denotes the electric charge, and $L$ signifies the parameter responsible for Lorentz violation. The metric function $A(r)$ incorporates the contribution of $L$, which modifies the standard RN geometry. If $L=0$, this line element reduces   to the standard RN BH solution. When both  $L$ and $Q$ vanish, the structure takes the form of the standard Schwarzschild BH.

To investigate the dynamical behavior of bosonic and fermionic perturbations in the RN-bumblebee spacetime, where $L$ modifies the background geometry. We employ the NP formalism, which provides a convenient null tetrad description of curved spacetime perturbations. Within this framework, the Teukolsky master equation offers a unified frame for massless fields of arbitrary spin $(s=0, 1/2, 1, 3/2, 2)$, as given below \cite{chandrasekhar1983}
\begin{equation}\label{eqn3}
\begin{split}
\Bigl\{&
\left[D-\epsilon(2s-1)+\epsilon^{*}
-2\rho s-\rho^{*}\right]
(\overline{\Delta}+\mu-2\gamma s) \\
&-
\left[\delta-\beta(2s-1)-\alpha^{*}
-2\tau s+\pi^{*}\right]
(\overline{\delta}+\pi-2\alpha s) \\
&-(2s-1)(s-1)\Psi_{2}
\Bigr\}\phi_{s}=0.
\end{split}
\end{equation}
and 
 \begin{equation}\label{eqn4}
\begin{split}
\Bigl\{&
[\overline{\Delta}+\gamma(2s-1)-\gamma^*+2\mu s
+\mu^*(D-\rho+2\epsilon s)] \\
&-[\overline{\delta}+\alpha(2s-1)+2\pi s+\beta^*-\tau^*]
(\delta-\tau+2\beta s) \\
&-(2s-1)(s-1)\Psi_2
\Bigr\}\phi_{-s}=0 .
\end{split}
\end{equation}
 Equations \eqref{eqn3} and \eqref{eqn4} describe the evolution of massless perturbation fields with spin $s=0, 1/2, 1, 3/2$ and $2$ in the RN-bumblebee spacetime. To separate the variables, we assume the wave functions in Eqs. \eqref{eqn3} and  \eqref{eqn4} that have the harmonic time and azimuthal dependence in the form $e^{i(\omega t +m \phi)}$ i.e,
 \begin{align} \label{eqn 5}
 & \phi_s = R_{+s}(r)A_{+s}(\theta)e^{i(\omega t +m \phi)} 
 \end{align}
 and
 \begin{align}
\label{eqn 6}
& \phi_{-s} = R_{-s}(r)A_{-s}(\theta)e^{-i(\omega t +m \phi)}.
 \end{align}

 Substituting above ansatzs into Eq. \eqref{eqn3} and introducing the transformation $P_{+s}= \Delta^s R_{+s}$ and $P_{-s}= r^{2s} R_{-s}$ the master equations reduce to the following decoupled radial and angular equations  
 \begin{align} \label{eqn 7}
 [ -2(2s-1)i\omega r+\Delta D_{1-s}D_{0}^+ ]P_{+s} = \lambda P_{+s},\\
  \label{eqb 8}
 L^+_{1-s}L_s A_{+s} = - \lambda A_{+s},
 \end{align}
 where $\lambda$ is a separation constant and $s$ can take the values of $0,1/2,1,3/2,2$.
 Similarly, separable solution of Eq. \eqref{eqn4} in terms of $P_{-s}$ and $A_{-s}$ satisfy the following equations
 \begin{align} \label{eqn 9}
 [2(2s-1)i\omega r + \Delta D^+_{1-s}D_{0}]P_{-s}= \lambda P_{-s}, \\ \label{eqn 10}
 L_{1-s}L_s^+ A_{-s}= - \lambda A_{-s}.
 \end{align}
 
  The regularity conditions imposed on the angular functions at $\theta=0$ and $\theta=\pi$ determine the allowed values of $\lambda$.

For the RN-bumblebee BH spacetime, the separation constant $\lambda$ can be determined from the regularity conditions imposed on the angular functions. Since $L$ modifies only the radial part of the spacetime geometry the explicit form of $\lambda$ remains identical to the standard case. For bosonic and fermion perturbations, separation constant can be written as
\begin{align*}
\lambda= (l+|s|)(l-|s|+l), l=|s|,|s|+1...
\end{align*}
and
\begin{align*}
\lambda= (j+|s|)(j-|s|+l), j=|s|,|s|+1...,
\end{align*}
 where the parameters $l$ and $j$ denote the orbital and total angular momentum, respectively.
  Defining the tortoise coordinate transformation through
 $dr_*=\frac{r^2\sqrt{1+L}}{\Delta} dr$ the differential operators $D_0$ and $D_0^+$ are rewritten as
 \begin{align} \label{eqn 11}
 D_0=\dfrac{ \sqrt{1+L} ~r^2}{\Delta}\Lambda_+ \hspace{0.5cm} \rm {and}   \hspace{0.5cm} D_0^+= \dfrac{\sqrt{1+L}  ~r^2}{\Delta}\Lambda_-,
 \end{align} 
 where  $\Lambda_{\pm}=\dfrac{d}{dr_*} \pm i\omega$. By further performing the variable substitution
  $Y=r^{1-2s}P_+$, Eq. \eqref{eqn 7} gives
 \begin{equation}\label{eqn12}
\begin{split}
&\frac{(1+L)r^{2s+3}}{\Delta}
\Biggl\{
\Lambda^2Y
+\left[
\frac{d}{dr_*}
\ln\!\left(\frac{r^{4s}}{\Delta^s}\right)
\right]\Lambda_-Y
\Biggr\}
\\
&\qquad
+\Biggl[
\Delta^s\frac{d}{dr}
\left(
\frac{1}{\Delta^{s-1}}
\frac{d}{dr}r^{2s-1}
\right)
-r^{2s-1}\lambda
\Biggr]Y=0,
\end{split}
\end{equation}
 where
 \begin{align}\label{eqn 13}
  \Lambda^2=\dfrac{d^2}{d(r_*)^2}+\omega^2.
 \end{align}
 
   After some algebraic simplification, Eq. \eqref{eqn12} reduces to a Schrödinger-like wave equation of the form
 \begin{align} \label{eqn 14}
 \Lambda^2Y+P \Lambda_- Y- IY=0 ,
 \end{align}
  where the coefficient functions $P$ and $I$ are given by
\begin{align} \label{eqn 15}
P= \dfrac{d}{dr_*}\ln \dfrac{r^{4s}}{\Delta^s}
\end{align}
 and 
\begin{align} \label{eqn 16}
I=\dfrac{(1+L)\Delta}{r^4}\left[ \lambda-(2s-1)(s-1)(\dfrac{2\Delta}{r^2}-\dfrac{\Delta'}{r})\right] .
\end{align}
To cast the perturbation equation into a standard one-dimensional wave-like form, we introduce a radial function $Z$. Now Eq. \eqref{eqn 13} can be transformed into of the form
\begin{align} \label{eqn 17}
\Lambda^2 W= VW,
\end{align}
where $V$ represents potential. 
Following the Chandrasekhar transformation approach, the functions $Y$ and  $Z$ can be related through 
\begin{align}\label{eqn 18}
Y=\xi \Lambda_+ \Lambda_+ W + U \Lambda_+ W.
\end{align} 
Here $\xi$ and $U$ denote unknown functions of $r_*$. We derive the following coupled system of equations
\begin{align} \label{eqn 19}
\chi=\xi V +\dfrac{dT}{dr_*}, \\ \label{eqn 20}
\dfrac{d}{dr_*}\left( \dfrac{r^4s}{\Delta^s}\chi\right)=\dfrac{r^4s}{\Delta^s}\left(IT- 2i\omega\chi \right)+\beta , \\ \label{eqn 21}
\chi\left( \chi-\dfrac{dT}{dr_*}\right)+\dfrac{\Delta^s}{r^{4s}}
\beta T=\dfrac{\Delta^s}{r^{4s}}K,\\ \label{eqn 22}
\chi V - I \xi V = \dfrac{\Delta^s}{r^{4s}}\dfrac{d \beta}{dr_*},
\end{align}
 where $K$ is a constant and $\chi$, $T$, and $\beta$ operate as functions dependent on  $r_*$.
To proceed further, we seek a consistent set of functions $\xi$, $\beta$, $T$ and $V$  to hold Eqs. {\eqref{eqn 19}-\eqref{eqn 22}}. Following the standard perturbation framework, the unknown terms are decomposed appropriately so that the resulting equations can be separated into  real and imaginary parts independently. Accordingly, without loss of generality, the functions $T, \beta$ and the constant $\kappa$ are assumed to take the form
\begin{equation}\label{eqn23}
\begin{aligned}
T=T_1(r_*)+2i\omega h(s),\qquad
\beta=\beta_1(r_*)+2i\omega\beta_2,\\
K=\kappa_1+2i\omega\kappa_2,
\end{aligned}
\end{equation}
 where $\beta_2, \kappa_1, \kappa_2$ represent constants and $h(s)$ is the function of $s$. We choose $h(s)$ as
 \begin{align} \label{eqn 24}
 h(s)=\dfrac{1}{6}s(2s-1)(6s^2- 23s+ 23), \hspace{0.2 cm} \rm{for} \hspace{0.2 cm}
 s= 0, \dfrac{1}{2}, 1,\dfrac{3}{2}, 2.
 \end{align}
 By substituting Eq. \eqref{eqn23} in Eqs. \eqref{eqn 20} and \eqref{eqn 21}, we obtain 
 \begin{align}\label{eqn 25}
 &\chi = hI+ \dfrac{\Delta^s}{r^{4s}}\beta_2,\\ \label{eqn 26}
 &\dfrac{d}{dr_*}\left( \dfrac{r^{4s}}{\Delta^s}\chi\right) = \dfrac{r^{4s}}{\Delta^s} IT_1+\beta_1
 \end{align}
 and
 \begin{align}\label{eqn 27}
 &\beta_1 h +\beta_2T_1 = \kappa_2,\\ \label{eqn 28}
 &\chi^2- \chi\dfrac{dT}{dr_*}+\dfrac{\Delta^s}{r^{4s}}\beta_1 T_1=\dfrac{\Delta^s}{r^{4s}}\kappa,
\end{align}
where $\kappa= \kappa_1 + 4\omega^2 h \beta_2$. 
  
  Substituting Eqs. \eqref{eqn 25}, \eqref{eqn 27} into  \eqref{eqn 26}, we get
  \begin{align} \label{eqn 29}
  T_1= \dfrac{h^2F_{,r_*}-\kappa_2}{Fh-\beta_2}.
  \end{align}
We also define $F=\dfrac{r^{4s}I}{\Delta^s}$ and $F_{,r_*}$ denotes differentiation of $F$ w.r.t. $r_*$.
 Then Eq. \eqref{eqn 28} can be written as 
  \begin{align} \label{eqn 30}
  \dfrac{\Delta^s}{r^{4s}}\left( hF +\beta_2 \right)^2-h^2\dfrac{(hF+\beta_2)F_{,r_*,r_*}}{hF-\beta_2} + \nonumber\\
  \dfrac{(f^4F^2_{,r_*}-\kappa_2^2)F}{(hF-\beta_2)^2}=\kappa.
  \end{align}
Since $\kappa_2$ appears as $\kappa_2^2$ in Eq. \eqref{eqn 30}, 
two branches of solution exist corresponding to 
$+\kappa_2$ and $-\kappa_2$ respectively, giving rise 
to two families of effective potentials $V^{(+)}$ and 
$V^{(-)}$. For the RN BHs in bumblebee gravity, 
substituting the metric functions and the 
LSB parameter $L$, the parameters 
$\beta_2$, $\kappa$, and $\kappa_2$ take the following forms
\begin{align}\label{eqn. beta}
\beta_2 &= -\frac{1}{3}(s-2)(2s-3)(4s-1)\lambda, \\
\kappa   &= \frac{1}{6} \lambda(\lambda+s)(s-1)(2s-3) (2s-1)(5s-8)
            , \\
\kappa_2 &= s(s-1)\Biggl[
\frac{8}{3}\lambda(2-s)
\sqrt{s-\frac{1}{2}+\lambda}
\nonumber\\
&\qquad +(2s-3)(2s-1)
\left(
M-\frac{2(1+L)}{2+L}\frac{Q^2}{r}
\right)
\Biggr],
\end{align}

where $\lambda = l(l+1)-|s|(|s|-1)$ is the spin-weighted 
spheroidal eigenvalue. 

  It is worth noting that in the Schwarzschild case 
\cite{Fu-Wen Shua2005}, the parameter $\kappa_2=6M$ 
is simply a mass-dependent constant. However, in the 
RN-bumblebee background, $\kappa_2$ becomes explicitly 
$r$-dependent through the charge correction term 
$\tfrac{2(1+L)}{2+L}\tfrac{Q^2}{r}$, representing 
a nontrivial coupling between the gravitational 
perturbation and the background electromagnetic and 
Lorentz-violating fields. In the limit $Q\rightarrow 0$, 
one correctly recovers $\kappa_2 = 6M$, serving as a 
consistency check of the derivation. Furthermore, in 
the simultaneous limit $Q\rightarrow 0$ and 
$L\rightarrow 0$, all effective potentials reduce to 
their standard Schwar- \\ zschild forms, providing a 
further verification of our results.
Substituting the values of $\kappa_2$,  $\beta_2$ and $\kappa$ into Eq. \eqref{eqn 22}, the solution of $V$ can be obtained as
  \begin{align}
 V= \dfrac{\Delta^s}{r^{4s}}F-\dfrac{(hF-\beta_2)hF_{,r_*,r_*}-h^2F^2_{,r_*}}{(hF-\beta_2)^2} +\dfrac{\kappa_2 F_{,r_*}}{(hF-\beta_2)^2}.
  \end{align}
   After simplyfying Eq. \eqref{eqn 7} to a one dimensional wave-equation, we get
   \begin{align}\label{waveeq}
   \dfrac{d^2 Z}{dr_*^2}+\omega^2W = VW.
   \end{align}
 The effective potential can be obtained as 
 \begin{align}
 V=\dfrac{\Delta}{r^{4s}}F- \dfrac{(hF-\beta_2)hF_{,r_* r_*}- h^2 F^2_{,r_*}+\kappa_2F_{,r_*}}{(hF-\beta_2)^2}.
 \end{align}\\

For scalar field ($s=0$), we get the following constants as 
\begin{align}
\beta_2\big|_{s=0} = -2\lambda, \qquad
\kappa\big|_{s=0}  = -4\lambda^2, \qquad
\kappa_2\big|_{s=0} = 0.
\end{align}

Since $\kappa_2=0$, the two branches coincide and 
yield a unique effective potential as

\begin{align}
V_0(r) = \frac{\Delta}{(1+L)r^4}
\left[\lambda - (2s-1)(s-1)\left(
\frac{2\Delta}{r^2} - \frac{\Delta'}{r}\right)\right].
\end{align}

For Dirac field ($s=1/2$), the constants terms are given by
\begin{align}
\beta_2\big|_{s=1/2}
= -\lambda,
\qquad
\kappa\big|_{s=1/2} = 0,
\qquad
\kappa_2\big|_{s=1/2} = 0,
\end{align}

which gives effective potential for the Dirac field as
\begin{align}
V_{1/2}(r) = \frac{\Delta^s}{r^{4s}}F 
- \frac{4F_1}{9},
\end{align}

where $\lambda_{1/2} = \left(l+\tfrac{1}{2}\right)^2$ 
for half-integer spin.\\

For electromagnetic field ($s=1$), we calculate $\beta_2$, $\kappa$ and $\kappa_2$ as

\begin{align}
\beta_2\big|_{s=1} = -\lambda, \qquad
\kappa\big|_{s=1}  = 0,        \qquad
\kappa_2\big|_{s=1} = 0.
\end{align}
Again $\kappa_2=0$,  we get the unique effective 
potential as
\begin{align}
V_1(r) = \frac{2\Delta}{r^4(1+L)}.
\end{align}

 For gravitino field ($s=3/2$), the values of $\beta_2$, $\kappa$ and $\kappa_2$  are

\begin{align}
\beta_2\big|_{s=3/2} 
= 0,
\qquad
\kappa\big|_{s=3/2}  
= 0,
\qquad
\kappa_2\big|_{s=3/2} = 0.
\end{align}

Since all three parameters vanish, the gravitino 
effective potential simplifies to

\begin{align}
V_{3/2}(r) = \frac{\Delta^s}{r^{4s}}F 
- \frac{(fF - \beta)fF_2 
- f^2F_1^2 + \kappa_2 F_1}
{(fF - \beta)^2},
\end{align}

where, $\lambda_{3/2} = \left(l+\frac{3}{2}\right)
\left(l-\frac{1}{2}\right)$.\\

For gravitational field ($s=2$), we derive the values of $\beta_2$, $\kappa$ and $\kappa_2$ as

\begin{align}
&\beta_2\big|_{s=2} = 0, \qquad
\kappa\big|_{s=2}  = -\lambda(\lambda+2), \nonumber\\ 
\qquad
&\kappa_2\big|_{s=2} 
= 6\left(M-\frac{2(1+L)}{2+L}\frac{Q^2}{r}\right).
\end{align}

Since $\kappa_2\neq 0$, two distinct effective 
potentials arise

\begin{align}
V_2^{(\pm)}(r) = \frac{\Delta^s}{r^{4s}}F 
- \frac{(fF - \beta)fF_2 
- f^2F_1^2 \pm \kappa_2 F_1}
{(fF - \beta)^2},
\end{align}
 where $\lambda=l(l+1)-2$, $\Delta = r^2 A(r)$, $f = A(r)$, $F$ is 
the tortoise-coordinate dependent function 
defined above, $F_1 = dF/dr_*$, 
$F_2 = d^2F/dr_*^2$, and $\beta \equiv \beta_2$ 
is given in Eq. \eqref{eqn. beta}. The tortoise coordinate 
$r_*$ is related to $r$ through 
${dr_*}/{dr} = {\sqrt{1+L}}/{A(r)}.$

\begin{figure}[!t]
\centering
\includegraphics[width=0.9\columnwidth]{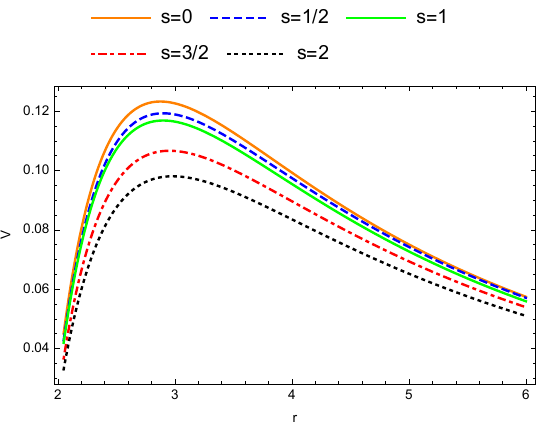}
\caption{Plot of the effective potential for different values of $s$ with
$M=1$, $L=0.3$, and $Q=0.3$.}
\label{fig:f1}
\end{figure}

\begin{figure*}[!t]
\centering

\subfloat[$s=0$]{\includegraphics[width=0.42\textwidth]{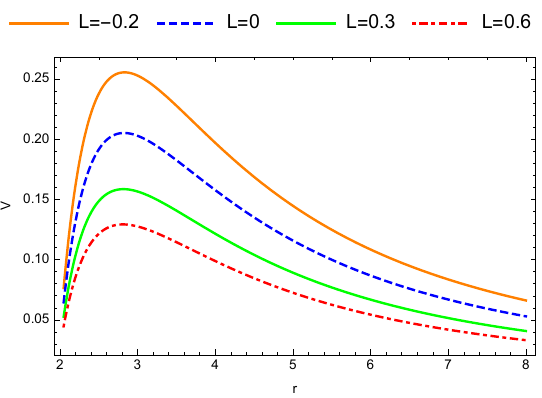}}
\hfill
\subfloat[$s=\frac12$]{\includegraphics[width=0.42\textwidth]{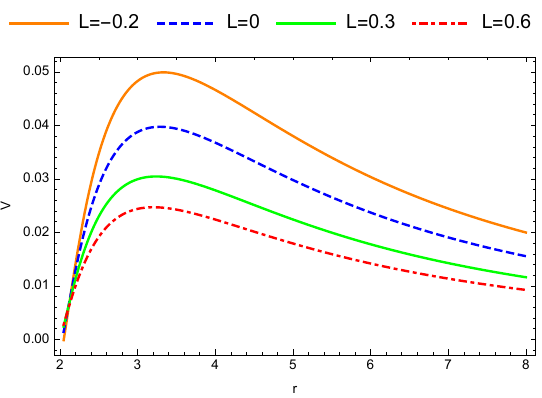}}

\vspace{3mm}

\subfloat[$s=1$]{\includegraphics[width=0.42\textwidth]{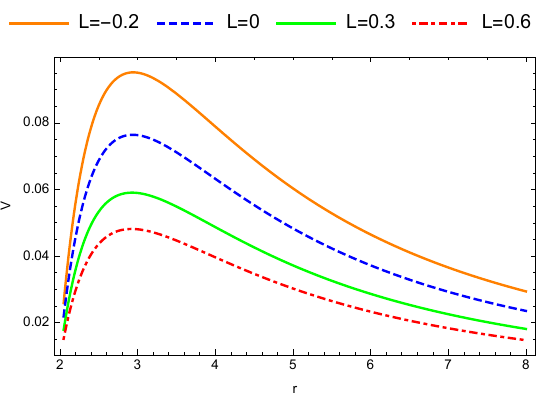}}
\hfill
\subfloat[$s=\frac32$]{\includegraphics[width=0.42\textwidth]{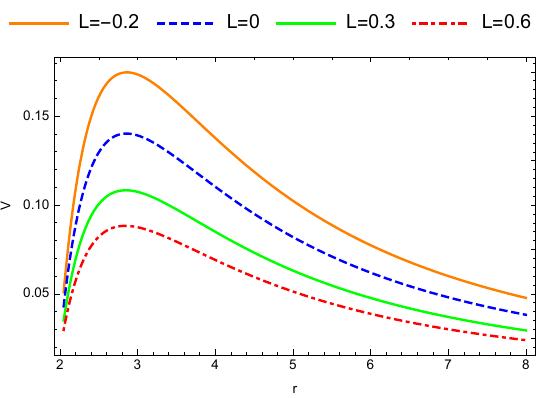}}

\vspace{3mm}

\subfloat[$s=2$]{\includegraphics[width=0.42\textwidth]{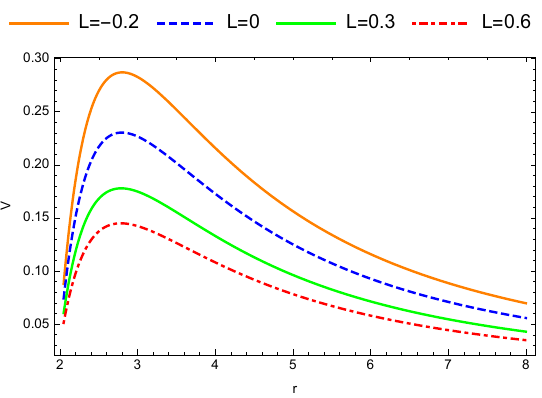}}

\caption{Plot of the effective potential for different values of $L$ with (a) $s=0$, (b) $s=\frac12$, (c) $s=1$, (d) $s=\frac32$, and (e) $s=2$.}
\label{fig:potential_L}
\end{figure*}

\begin{figure*}[!t]
\centering

\subfloat[$s=0$]{%
\includegraphics[width=0.42\textwidth]{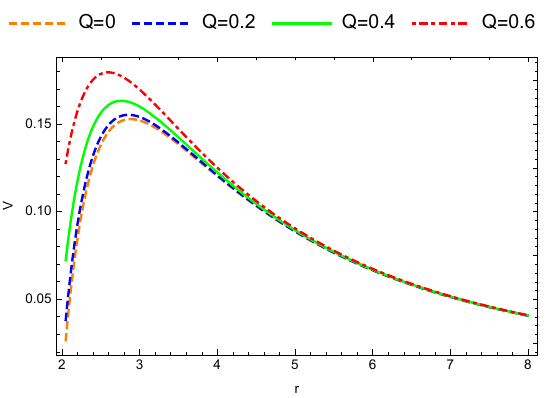}
\label{fig:f4a}}
\hfill
\subfloat[$s=\frac12$]{%
\includegraphics[width=0.42\textwidth]{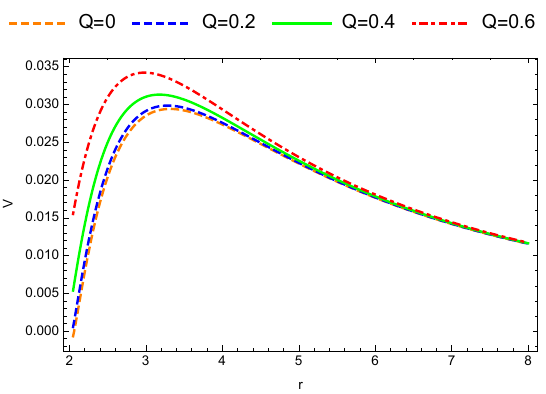}
\label{fig:f4b}}

\vspace{3mm}

\subfloat[$s=1$]{%
\includegraphics[width=0.42\textwidth]{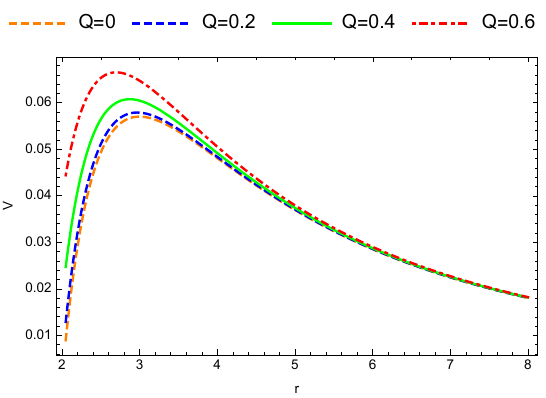}
\label{fig:f4c}}
\hfill
\subfloat[$s=\frac32$]{%
\includegraphics[width=0.42\textwidth]{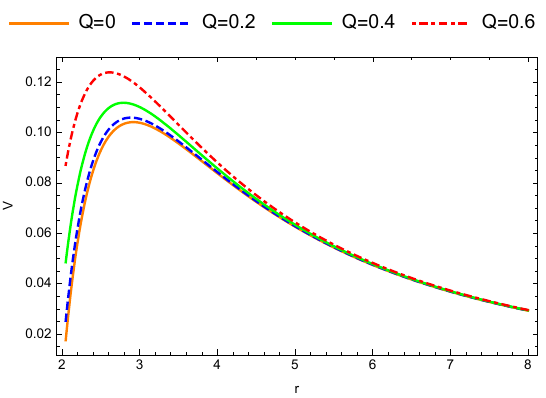}
\label{fig:f4d}}

\vspace{3mm}

\subfloat[$s=2$]{%
\includegraphics[width=0.42\textwidth]{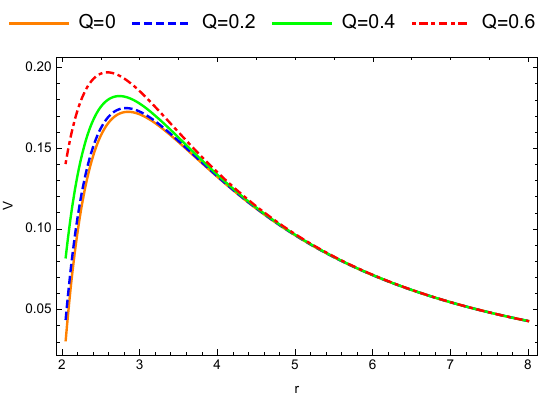}
\label{fig:f4e}}

\caption{Plot of the effective potential for different values of $Q$ with (a) $s=0$, (b) $s=\frac12$, (c) $s=1$, (d) $s=\frac32$, and (e) $s=2$.}
\label{fig:potential_Q}
\end{figure*}

Figure \ref{fig:f1} shows the variation of effective potential $V$ with different spin  (s= 0, 1/2, 1, 3/2, 2) values. The effective potential exhibits a positive potential barrier for all spin fields. The peak of the potential decreases with increasing spin, implying weaker confinement for higher-spin fields. Fermionic perturbations possess slightly higher potential barriers compared to bosonic perturbations.

Figures \ref{fig:potential_L}-\ref{fig:potential_Q} illustrate the behaviour of the effective potential derived using Teukolsky master equation within the NP formalism, which provides a unified framework for analyzing perturbations of different spin fields. Fig. \ref{fig:potential_L} represents the variation of  $V(r)$ for different values of the  $L$ corresponding to scalar, Dirac, electromagnetic, Rarita–Schwinger, and gravitational field perturbations in the RN-bumblebee BH spacetime, keeping all other parameters fixed.
It is observed that, for all spin values, the peak of the effective potential decreases with increasing values of the  parameter $L$. This indicates that the LSB parameter weakens the potential barrier experienced by the perturbing fields. Figure \ref{fig:potential_Q} describes the behavior of the effective potential $V$ for different values of charge $Q$ for different spin fields. It is evident that increasing the value of $Q$ raises the peak of the effective potential uniformly across all  field perturbations
\section{QNMs frequencies for different spin fields}

In this section, we investigate the QNMs  of massless perturbation fields of arbitrary spin  of the charged Lorentz-violating BH. To compute these QNM frequencies, we implement two   numerical approaches, namely the Padé-improved 6th-order WKB approximation and the improved AIM. A comparison between the outcomes is subsequently provided, allowing us to evaluate the precision  and consistency of the calculated QNM spectra.
\subsection{WKB method}
The QNM frequencies are first computed using the WKB approximation, which provides an efficient semi-analytical method for solving black-hole perturbation equations. The WKB method is valid under the condition that the underlying effective potential possesses a solitary peak structure and the perturbations satisfy the QNM boundary conditions,
\begin{align}
\Psi(r_*) &\sim e^{-i\omega r_*}, \qquad r_* \rightarrow -\infty,\\
\Psi(r_*) &\sim e^{\,i\omega r_*}, \qquad r_* \rightarrow +\infty,
\end{align}

corresponding to purely ingoing waves at the event horizon and purely outgoing waves at spatial infinity. We consider the  6th-order WKB approximation method given by
\begin{eqnarray}
\dfrac{i(\omega^2-V_0)}{\sqrt{-2V''_0}}-\sum_{i=2}^6\Lambda_i=n+\dfrac{1}{2},
\end{eqnarray}
where $n$ is the overtone number. Here, $V_0$ and $V_0''$ denote the value of the effective potential and its corresponding second-order derivative calculated  with respect to $r_*$, respectively, with both values evaluated precisely at the maximum of the  potential. The quantities $\Lambda_i$ represent the higher-order correction terms of the WKB expansion and are functions of the higher-order derivatives of the effective potential. Their exact algebraic expressions are available in Refs.~\cite{S.Iyer1987,konoplya2003}. Although the higher-order WKB approximation provides accurate quasinormal frequencies for a wide range of black-hole potentials, its convergence may deteriorate for certain values of the multipole and overtone numbers. To enhance the convergence and improve the numerical accuracy of the WKB  series, we employ the Padé resummation technique proposed by Matyjasek and Opala \cite{MatyjasekOpala2017}. In the present work, the quasinormal frequencies are obtained from the Padé-improved 6th-order WKB approximation, which has been shown to provide significantly more accurate results than the conventional WKB expansion.

\subsection{AIM method}

Although the WKB approximation provides accurate estimates of the quasinormal frequencies, it is preferable to verify  the results using an alternative technique. For this purpose, we apply the AIM, which has been extensively used in black-hole perturbation studies and is known for its high numerical accuracy. The frequencies obtained through AIM are then compared with the WKB results and the agreement between the two methods serves as an important consistency check for the calculated spectra. To facilitate the numerical implementation of AIM, we introduce the new coordinate
transformation
\begin{align}
x=\frac{1}{r},
\end{align}
which maps the exterior region of the BH onto a finite interval. Under this transformation, Eq.~(\ref{waveeq}) becomes
\begin{equation}\label{AIMmaster}
\frac{u(x)^2}{1+L}\,\frac{d^2\Psi}{dx^2}
+\frac{u(x)~u'(x)}{1+L}\,\frac{d\Psi}{dx}
+\left[\omega^2-V(x)\right]\Psi=0,
\end{equation}

with
\begin{equation}
u(x)=x^2\left(1-2Mx+\frac{2(1+L)Q^2}{2+L}x^2\right).
\end{equation}

The event horizon $r_h$ is determined by the largest positive root of $A(r_h)=0$, and we define $x_h=1/r_h$. Near the horizon, the quasinormal boundary condition requires a purely ingoing wave,

\begin{equation}
\Psi \sim e^{-i\omega r_*},
\qquad r\rightarrow r_h.
\end{equation}

The associated surface gravity is

\begin{equation}
\kappa_h=
\frac{1}{2\sqrt{1+L}}
\left.
\frac{dA(r)}{dr}
\right|_{r=r_h}.
\end{equation}

To remove the singular behavior at the event horizon, the wave function is decomposed as

\begin{equation}
\Psi(x)
=e^{-i\omega r_*}
(x-x_h)^{-\frac{i\omega}{\kappa_h}}
 \chi(x),
\label{ansatz}
\end{equation}

where the function $\chi(x)$ is regular at $x=x_h$. Substituting Eq.~(\ref{ansatz}) into Eq.~(\ref{AIMmaster}) and simplifying yields

\begin{equation}
\chi''(x)=
\lambda_0(x)\chi'(x)
+
\chi(x) s_0(x),
\label{AIMform}
\end{equation}

which is the standard form required by the AIM. The coefficient  $\lambda_0(x)$ is identical for all perturbation spins, as it is determined solely by the background geometry and the adopted coordinate transformation. In contrast, the effective potential depends explicitly on the spin of the perturbing field, leading to different forms of the AIM coefficient  $s_0(x)$.   The explicit forms of  $\lambda_0(x)$ and
$s_0^{(s)}(x)$ used in the numerical implementation are provided in
Appendix A.

The AIM procedure is based on repeated differentiation of Eq.~(\ref{AIMform}), leading to

\begin{equation}
\chi^{(n+2)}=
\chi(x) s_n(x) +\lambda_n(x)\chi'(x),
\end{equation}

where the sequences $\lambda_n(x)$ and $s_n(x)$ satisfy the recursion relations

\begin{align}
\lambda_n(x)
&=
\lambda_{n-1}'(x)
+\lambda_0(x)\lambda_{n-1}(x)+s_{n-1}(x),
\\
s_n(x)
&=
s_{n-1}'(x)
+\lambda_{n-1}(x) s_0(x).
\end{align}

To improve numerical stability and efficiency, the improved AIM method is employed. Expanding $\lambda_n(x)$ and $s_n(x)$ around a suitable point $x=\bar{x}$,

\begin{align}
\lambda_n(x)
&=
\sum_{k=0}^{\infty}
c_n^{k}(x-\bar{x})^k,
\\
s_n(x)
&=
\sum_{k=0}^{\infty}
d_n^{k}(x-\bar{x})^k,
\end{align}

one obtains the recurrence relations

\begin{align}
c_n^{k}
&=
(k+1)c_{n-1}^{k+1}
+d_{n-1}^{k}
+
\sum_{m=0}^{k}
c_0^{m}
c_{n-1}^{k-m},
\\
d_n^{k}
&=
(k+1)d_{n-1}^{k+1}
+
\sum_{m=0}^{k}
d_0^{m}
c_{n-1}^{k-m}.
\end{align}

The quasinormal frequencies are determined by imposing the AIM termination condition

\begin{equation}
d_n^{0}c_{n-1}^{0}
- d_{n-1}^{0}c_n^{0}
=
0.
\label{termination}
\end{equation}

\begin{table*}[!t]
\centering
\caption{ Quasinormal frequencies for different perturbation fields for different values of $L$ and $Q$, computed using the 6th-order WKB method with Padé approximants and the AIM method for $\ell=2$, $n=0$. }
\label{tab QNM}

\textbf{(a) Scalar perturbation}

\medskip

\begin{tabular}{ccccccc}
\hline
$L$ & $Q$ 		&$\omega_{\rm WKB}$ & $\omega_{\rm AIM}$ &Relative Error\\
\hline
-0.1 & 0.1 		& $	0.499221 - 0.0850581i$ & $	0.499240 -0.0850522i$ 	&	0.00393\\
-0.1 & 0.3 		& $	0.504957 - 0.0857786	i$ & $	0.504975 -0.0857728i$ 	&	0.00368\\
-0.1 & 0.5 		& $	0.517576 - 0.0872186	i$ & $	0.517593 -0.0872129i$ 	&	0.00336\\ \hline
0 & 0.1 			& $	0.47364 - 0.0806979	i$ & $	0.473657 -0.0806923i$ 	&	0.00377\\
0 & 0.3 			& $	0.479396 - 0.0814195	i$ & $0.479413 -0.0814141	i$ 	&	0.00356\\
0 & 0.5 			& $	0.492133 - 0.0828601	i$ & $0.492149 -0.0828547	i$ 	&	0.00326\\ \hline
0.1 & 0.1 			& $	0.451629 - 0.0769465	i$ & $0.451646 -0.0769412	i$ 	&	0.00382\\
0.1 & 0.3 		& $	0.45739 - 0.0776674	i$ & $	0.457406 -0.0776622i$ 	&	0.00361\\
0.1 & 0.5 		& $	0.470206 - 0.0791049	i$ & $	0.470221 -0.0790998i$ 	&	0.00329\\ \hline
0.2 & 0.1 			& $		0.43243 - 0.0736742i$ & $	0.432446 -0.0736691i$ 	&	0.00384\\
0.2 & 0.3 		& $		0.438183 - 0.0743931i$ & $	0.438199 -0.0743881i$ 	&	0.00363\\
0.2 & 0.5 		& $	0.451046 - 0.0758247 	i$ & $	0.451060 -0.0758198i$ 	&	0.00330\\

\hline
\end{tabular}

\vspace{5mm}

\textbf{(b) Dirac perturbation}

\medskip

\begin{tabular}{ccccccc}
\hline
$L$ & $Q$ 		&$\omega_{\rm WKB}$ & $\omega_{\rm AIM}$ &Relative Error\\
\hline
-0.1 & 0.1 		& $	0.480459 - 0.101082	i$ & $	0.480461 -0.101078	i$ 	&	0.00089\\
-0.1 & 0.3 		& $	0.486904 - 0.101504		i$ & $	0.486906 -0.101500	i$ 	&	0.00087\\
-0.1 & 0.5 		& $	0.501019 - 0.102269		i$ & $	0.501021 -0.102265	i$ 	&	0.00082\\ \hline
0 & 0.1 			& $	0.481295 - 0.0959772	i$ & $	0.481296-0.0959743	i$ 	&	0.00062\\
0 & 0.3 			& $	0.488098 - 0.096398		i$ & $0.488099 -0.0963953		i$ 	&0.00057	\\
0 & 0.5 			& $	0.503078 - 0.097152	i$ & $	0.503079 -0.0971497	i$ 	&	0.00049\\ \hline
0.1 & 0.1 		& $	0.481954 - 0.0915738	i$ & $	0.481955 -0.0915717	i$ 	&	0.00046\\
0.1 & 0.3 		& $	0.489082 - 0.0919923	i$ & $	0.489083 -0.0919903	i$ 	&	0.00042\\
0.1 & 0.5 		& $	0.504856 - 0.0927349	i$ & $	0.504857 -0.0927333	i$ 	&	0.00035\\ \hline
0.2 & 0.1 		& $	0.482483 - 0.0877245		i$ & $	0.482484 -0.0877229	i$ 	&	0.00037\\
0.2 & 0.3 		& $		0.489909 - 0.0881402	i$ & $	0.489909 -0.0881387	i$ 	&	0.00030\\
0.2 & 0.5 		& $	0.506413 - 0.0888711	i$ & $	0.506414 -0.0888699	i$ 	&	0.00028\\

\hline
\end{tabular}

\vspace{5mm}

\textbf{(c) Electromagnetic perturbation}

\medskip

\begin{tabular}{ccccccc}
\hline
$L$ & $Q$ 		&$\omega_{\rm WKB}$ & $\omega_{\rm AIM}$ &Relative Error\\
\hline
-0.1 & 0.1 		& $	0.483144 - 0.100203i$ & $	0.483146 -0.100199i$ 	&	0.00089\\
-0.1 & 0.3 		& $	0.489738 - 0.100637	i$ & $	0.489740 -0.100634i$ 	&	0.00073\\
-0.1 & 0.5 		& $	0.504196 - 0.101429	i$ & $	0.504197 -0.101426i$ 	&	0.00061\\ \hline
0 & 0.1 			& $	0.458393 - 0.0950636	i$ & $0.458395 -0.0950601	i$ 	&0.00084	\\
0 & 0.3 			& $		0.465009 - 0.0954979i$ & $	0.465011 -0.0954946i$ 	&	0.00077\\
0 & 0.5 			& $	0.479599 - 0.0962824	i$ & $	0.479600 -0.0962795i$ 	&	0.00062\\ \hline
0.1 & 0.1 		& $	0.437097 - 0.0906421	i$ & $	0.437099 -0.0906387i$ 	&	0.00090\\
0.1 & 0.3 		& $	0.443718 - 0.0910753	i$ & $0.443720 -0.0910721	i$ 	&	0.00081\\
0.1 & 0.5 		& $	0.458394 - 0.0918512	i$ & $0.458395 -0.0918485	i$ 	&	0.00061\\ \hline
0.2 & 0.1 		& $		0.41852 - 0.0867854i$ & $	0.418522 -0.0867822i$ 	&	0.00095\\
0.2 & 0.3 		& $		0.425133 - 0.0872167i$ & $	0.425134 -0.0872136i$ 	&	0.00073\\
0.2 & 0.5 		& $	0.439858 - 0.0879831 	i$ & $	0.439859 -0.0879805i$ 	&	0.00060\\

\hline
\end{tabular}
\end{table*}
\begin{table*}[!t]
\centering
\ContinuedFloat
\centering
\textbf{(d) Rarita–Schwinger perturbation}

\medskip
\begin{tabular}{ccccccc}
\hline
$L$ & $Q$ 		&$\omega_{\rm WKB}$ & $\omega_{\rm AIM}$ &Relative Error\\
\hline
-0.1 & 0.1 		& $	0.410077 - 0.0986254	i$ & $	0.410081 -0.0986177	i$ 	&	0.00192\\
-0.1 & 0.3 		& $	0.415889 - 0.0990801		i$ & $	0.415893 -0.0990725	i$ 	&	0.00189\\
-0.1 & 0.5 		& $		0.428662 - 0.099917	i$ & $	0.428666 -0.0999100	i$ 	&	0.00177\\ \hline
0 & 0.1 			& $	0.412568 - 0.0937903	i$ & $	0.412571 -0.0937848	i$ 	&	0.00140\\
 0 & 0.3 			& $	0.418708 - 0.0942412		i$ & $	0.418711 -0.0942358	i$ 	&	0.00136\\
0 & 0.5 			& $ 0.432273 - 0.0950634 	i$ & $	0.432275 -0.0950585	i$ 	&	0.00117\\ \hline
0.1 & 0.1 		& $	0.41464 - 0.0896036 	i$ & $	0.414642 -0.0895996	i$ 	&	0.00099\\
0.1 & 0.3 		& $	0.421077 - 0.0900505	i$ & $	0.421080 -0.0900465	i$ 	&	0.00111\\
0.1 & 0.5 		& $	0.435369 - 0.0908577 i$ & $	0.435371 -0.0908542	i$ 	&	0.00088\\ \hline
0.2 & 0.1 		& $		0.416394 - 0.0859325	i$ & $	0.416395 -0.0859295	i$ 	&	0.00073\\
0.2 & 0.3 		& $		0.423102 - 0.086375	i$ & $	0.423104 -0.086372	i$ 	&	0.00069\\
0.2 & 0.5 		& $	0.438064 - 0.0871674	i$ & $		0.438065 -0.0871648i$ 	&	0.00058\\

\hline
\end{tabular}
\vspace{5mm}

\textbf{(e) Gravitational perturbation}

\medskip
\begin{tabular}{ccccccc}
\hline
$L$ & $Q$ 		&$\omega_{\rm WKB}$ & $\omega_{\rm AIM}$ &Relative Error\\
\hline
-0.1 & 0.1 		& $	 0.972506 - 0.0986492	i$ & $		0.972505 -0.0986485	i$ 	&	0.00012\\
-0.1 & 0.3 		& $	 0.983091 - 0.098993	i$ & $	0.983091 -0.0989933		i$ 	&	0.00003\\
-0.1 & 0.5 		& $	 1.00629 - 0.099598	i$ & $	1.00629 -0.0995977		i$ 	&	0.00003\\ \hline
0 & 0.1 			& $	 0.97727 - 0.0937471	i$ & $	0.977271 -0.0937466		i$ 	&	0.00010\\
0 & 0.3 			& $	0.988489 - 0.09409	i$ & $	0.988489 -0.0940897		i$ 	&	0.00003\\
0 & 0.5 			& $	1.01321 - 0.0946863	i$ & $	1.01321 -0.0946853		i$ 	&	0.00010\\ \hline
0.1 & 0.1 		& $	0.981333 - 0.0895142	i$ & $	0.981334 -0.0895148		i$ 	&	0.00010\\
0.1 & 0.3 		& $	0.99313 - 0.0898556	i$ & $	0.993129 -0.0898541		i$ 	&	0.00016\\
0.1 & 0.5 		& $	1.01925 - 0.0904422 	i$ & $	1.01925 -0.0904420		i$ 	&	0.00002\\ \hline
0.2 & 0.1 		& $	0.984851 - 0.0858106	i$ & $	0.984849 -0.0858109		i$ 	&	0.00020\\
0.2 & 0.3 		& $	0.997175 - 0.0861499	i$ & $	0.997175 -0.0861515		i$ 	&	0.00016\\
0.2 & 0.5 		& $	1.02458 - 0.0867264	i$ & $	1.02458 -0.0867259		i$ 	&	0.00005\\

\hline
\end{tabular}
\end{table*}
\addtocounter{table}{1}

\subsection{Comparison of WKB and AIM Results}

Table \ref{tab QNM} summarize the fundamental quasinormal frequencies  obtained using the Pad\'e-improved WKB method and the improved AIM for scalar, Dirac, electromagnetic, Rarita--Schwinger and gravitational  perturbations. The two independent numerical approaches exhibit excellent agreement for all values of the  $L$ and $Q$.

For every perturbation sector, the discrepancies between the WKB and AIM frequencies appear only in the sixth or seventh decimal place. Such close agreement demonstrates the reliability of the derived effective potentials and the numerical stability of the calculations, providing a strong validation of the computed quasinormal spectra.

The influence of $L$ is clearly reflected in the quasinormal frequencies. For scalar and electromagnetic perturbations, increasing $L$ leads to a systematic decrease in both the oscillation frequency and the damping rate. Consequently, these modes oscillate more slowly and decay more gradually. A similar reduction in the damping rate is also observed for the Dirac, Rarita--Schwinger, and gravitational perturbations.

In contrast, the electric charge $Q$ exhibits the opposite trend. For fixed $L$, both the real and imaginary parts of the quasinormal frequencies generally increase with increasing charge, indicating higher oscillation frequencies accompanied by slightly faster damping. This behavior is consistently observed for all perturbation spins considered.

A comparison among the different perturbing fields further reveals a clear spin dependence of the spectrum. The gravitational perturbations  possess the largest oscillation frequencies, whereas the scalar, electromagnetic, Dirac, and Rarita--Schwinger fields exhibit comparatively smaller values of $\mathrm{Re}(\omega)$. These differences originate from the distinct effective potentials governing each perturbation sector.

Overall, the excellent agreement between the Pad\'e-improved WKB and AIM results demonstrates the reliability of the calculated quasinormal frequencies. The relative errors remain extremely small throughout the parameter space considered, further confirming the consistency of the two numerical approaches and the validity of the computed quasinormal spectra.

\section{Observational Prospects of QNMs in Bumblebee Gravity}
The detection of BH modes through gravitational wave observations provides a powerful way to test the gravity in the strong-field regime. In particular, deviations from general relativity, such as Lorentz symmetry breaking in bumblebee gravity, can leave observable imprints on the quasinormal spectrum of BHs. Motivated by the recent success of gravitational wave detectors, it is therefore important to examine whether the QNMs of the RN BHs in bumblebee gravity fall within the sensitivity bands of current and future observatories such as  Virgo, LIGO and LISA.

 To assess the observational relevance of the computed QNMs, the dimensionless frequencies obtained in the previous sections must be converted into physical units. Following the standard convention \cite{Ferrari2008}, we consider a BH of mass
\begin{equation}
M=\hat{\eta}M_\odot,
\end{equation}
where $\hat{\eta}$ is the BH mass scaled in terms of solar mass units  $M_\odot$, with $M_\odot=1.48\times10^{5}\,\mathrm{cm}$ in geometrized units. The real and imaginary parts of the calculated dimensionless quasinormal frequency can be mapped directly to the actual physical vibration frequency and characteristic damping timescale using the following relationships
\begin{align}
\hat{f}
&=\frac{c\,M\,\omega_R}{2\pi \hat{\eta}M_\odot}\,\mathrm{kHz}
=\frac{32.36\,M\,\omega_R}{\hat{\eta}}\,\mathrm{kHz},\\
\tau
&=\frac{\hat{\eta}M_\odot}{M\,\omega_I\,c}\,\mathrm{s}
=\frac{0.4937\times10^{-5}\,\hat{\eta}}
{M\,\omega_I}\,\mathrm{s},
\end{align}
where $\omega_R$ and $\omega_I$ denote the real and imaginary parts of the dimensionless quasinormal frequency, respectively.

These relations enable a direct comparison between the predicted QNMs frequencies and the operating frequency bands of present and future gravitational-wave detectors. Ground-based interferometers, including LIGO, Virgo, and KAGRA, are primarily sensitive to frequencies ranging from approximately $10$ Hz to $10^{3}$ Hz, making them suitable for observing the ringdown signals of stellar-mass BHs. On the other hand, the space-based LISA is designed to probe much lower frequencies, roughly between $10^{-4}$ Hz and $1$ Hz, allowing the detection of ringdown signals from intermediate and supermassive BHs.

\begin{table*}[t]
\centering
\caption{Detectable black-hole mass ranges corresponding to the scalar QNM ($\ell=2$, $n=0$). }
\label{tab:detectability0}
\begin{tabular}{c c c c c c}
\hline
\multicolumn{6}{c}{(a) Fixed $Q=0.3$, varying $L$}\\
\hline
$L$ & $\omega_R$ &
$M_{\min}^{\rm LIGO}(M_\odot)$ &
$M_{\max}^{\rm LIGO}(M_\odot)$ &
$M_{\min}^{\rm LISA}(M_\odot)$ &
$M_{\max}^{\rm LISA}(M_\odot)$\\
\hline
$-0.1$ & 0.504957 &13.5749& 1357.49& 16289.9& 1.62899 $\times 10^8$ \\
$0.0$  & 0.479396 & 12.8878 & 1288.78 & 15465.3 & 1.54653 $\times 10^8$ \\
$0.1$  & 0.457390 & 12.2962 & 1229.62 & 14755.4 & 1.47554$\times 10^8$ \\
\hline
\multicolumn{6}{c}{}\\[-2mm]
\hline
\multicolumn{6}{c}{(b) Fixed $L=0.1$, varying $Q$}\\
\hline
$Q$ & $\omega_R$ &
$M_{\min}^{\rm LIGO}(M_\odot)$ &
$M_{\max}^{\rm LIGO}(M_\odot)$ &
$M_{\min}^{\rm LISA}(M_\odot)$ &
$M_{\max}^{\rm LISA}(M_\odot)$\\
\hline
$0.1$ & 0.451629 & 12.1413 & 1214.13 & 14569.6 & 1.45696$\times 10^8$ \\
$0.3$ & 0.457390 & 12.2962 & 1229.62 & 14755.4 & 1.47554 $\times 10^8$ \\
$0.5$ & 0.470206 & 12.6407 & 1264.07 & 15168.8& 1.51688$\times 10^8$ \\
\hline
\end{tabular}
\end{table*}

\begin{table*}[t]
\centering
\caption{Detectable black-hole mass ranges corresponding to the  QNM  for $s=\frac{1}{2}$ ($\ell=2$, $n=0$). }
\label{tab:detectability12}
\begin{tabular}{c c c c c c}
\hline
\multicolumn{6}{c}{(a) Fixed $Q=0.3$, varying $L$}\\
\hline
$L$ & $\omega_R$ &
$M_{\min}^{\rm LIGO}(M_\odot)$ &
$M_{\max}^{\rm LIGO}(M_\odot)$ &
$M_{\min}^{\rm LISA}(M_\odot)$ &
$M_{\max}^{\rm LISA}(M_\odot)$\\
\hline
 -0.1 & 0.486904 & 13.0896 & 1308.96 & 15707.5 & 1.57075$\times 10^8$ \\
 0. & 0.488098 & 13.1217 & 1312.17 & 15746. & 1.5746$\times 10^8$ \\
 0.1 & 0.489082 & 13.1482 & 1314.82 & 15777.8 & 1.57778$\times 10^8$ \\
\hline
\multicolumn{6}{c}{}\\[-2mm]
\hline
\multicolumn{6}{c}{(b) Fixed $L=0.1$, varying $Q$}\\
\hline
$Q$ & $\omega_R$ &
$M_{\min}^{\rm LIGO}(M_\odot)$ &
$M_{\max}^{\rm LIGO}(M_\odot)$ &
$M_{\min}^{\rm LISA}(M_\odot)$ &
$M_{\max}^{\rm LISA}(M_\odot)$\\
\hline
 0.1 & 0.481954 & 12.9565 & 1295.65 & 15547.8 & 1.55478$\times 10^8$ \\
 0.3 & 0.489082 & 13.1482 & 1314.82 & 15777.8 & 1.57778$\times 10^8$ \\
 0.5 & 0.504856 & 13.5722 & 1357.22 & 16286.7 & 1.62867$\times 10^8$ \\
\hline
\end{tabular}
\end{table*}

\begin{table*}[t]
\centering
\caption{Detectable black-hole mass ranges corresponding to the  QNM  for $s=1$ ($\ell=2$, $n=0$). }
\label{tab:detectability1}
\begin{tabular}{c c c c c c}
\hline
\multicolumn{6}{c}{(a) Fixed $Q=0.3$, varying $L$}\\
\hline
$L$ & $\omega_R$ &
$M_{\min}^{\rm LIGO}(M_\odot)$ &
$M_{\max}^{\rm LIGO}(M_\odot)$ &
$M_{\min}^{\rm LISA}(M_\odot)$ &
$M_{\max}^{\rm LISA}(M_\odot)$\\
\hline
-0.1 & 0.489738 & 13.1658 & 1316.58 & 15798.9 & 1.57989$\times 10^8$ \\
 0. & 0.465009 & 12.501 & 1250.1 & 15001.2 & 1.50012$\times 10^8$ \\
 0.1 & 0.443718 & 11.9286 & 1192.86 & 14314.3 & 1.43143$\times 10^8$ \\
\hline
\multicolumn{6}{c}{}\\[-2mm]
\hline
\multicolumn{6}{c}{(b) Fixed $L=0.1$, varying $Q$}\\
\hline
$Q$ & $\omega_R$ &
$M_{\min}^{\rm LIGO}(M_\odot)$ &
$M_{\max}^{\rm LIGO}(M_\odot)$ &
$M_{\min}^{\rm LISA}(M_\odot)$ &
$M_{\max}^{\rm LISA}(M_\odot)$\\
\hline
 0.1 & 0.437097 & 11.7506 & 1175.06 & 14100.7 & 1.41007$\times 10^8$ \\
 0.3 & 0.443718 & 11.9286 & 1192.86 & 14314.3 & 1.43143$\times 10^8$ \\
 0.5 & 0.458394 & 12.3232 & 1232.32 & 14787.8 & 1.47878$\times 10^8$ \\
\hline
\end{tabular}
\end{table*}

\begin{table*}[t] 
\centering
\caption{Detectable black-hole mass ranges corresponding to the  QNM  for $s=\frac{3}{2}$ ($\ell=2$, $n=0$). }
\label{tab:detectability32}
\begin{tabular}{c c c c c c}
\hline
\multicolumn{6}{c}{(a) Fixed $Q=0.3$, varying $L$}\\
\hline
$L$ & $\omega_R$ &
$M_{\min}^{\rm LIGO}(M_\odot)$ &
$M_{\max}^{\rm LIGO}(M_\odot)$ &
$M_{\min}^{\rm LISA}(M_\odot)$ &
$M_{\max}^{\rm LISA}(M_\odot)$\\
\hline
-0.1 & 0.415889 & 11.1805 & 1118.05 & 13416.6 & 1.34166$\times 10^8$ \\
 0. & 0.418708 & 11.2563 & 1125.63 & 13507.5 & 1.35075$\times 10^8$ \\
 0.1 & 0.421077 & 11.32 & 1132. & 13583.9 & 1.35839$\times 10^8$ \\
\hline
\multicolumn{6}{c}{}\\[-2mm]
\hline
\multicolumn{6}{c}{(b) Fixed $L=0.1$, varying $Q$}\\
\hline
$Q$ & $\omega_R$ &
$M_{\min}^{\rm LIGO}(M_\odot)$ &
$M_{\max}^{\rm LIGO}(M_\odot)$ &
$M_{\min}^{\rm LISA}(M_\odot)$ &
$M_{\max}^{\rm LISA}(M_\odot)$\\
\hline
  0.1 & 0.41464 & 11.1469 & 1114.69 & 13376.3 & 1.33763$\times 10^8$ \\
 0.3 & 0.421077 & 11.32 & 1132. & 13583.9 & 1.35839$\times 10^8$ \\
 0.5 & 0.435369 & 11.7042 & 1170.42 & 14045. & 1.4045$\times 10^8$ \\
\hline
\end{tabular}
\end{table*}

\begin{table*}[t]
\centering
\caption{Detectable black-hole mass ranges corresponding to the  QNM  for $s=2$ ($\ell=2$, $n=0$). }
\label{tab:detectability2}
\begin{tabular}{c c c c c c}
\hline
\multicolumn{6}{c}{(a) Fixed $Q=0.3$, varying $L$}\\
\hline
$L$ & $\omega_R$ &
$M_{\min}^{\rm LIGO}(M_\odot)$ &
$M_{\max}^{\rm LIGO}(M_\odot)$ &
$M_{\min}^{\rm LISA}(M_\odot)$ &
$M_{\max}^{\rm LISA}(M_\odot)$\\
\hline
 -0.1 & 0.983091 & 26.4288 & 2642.88 & 31714.5 & 3.17145$\times 10^8$ \\
 0. & 0.988489 & 26.5739 & 2657.39 & 31888.7 & 3.18887$\times 10^8$ \\
 0.1 & 0.99313 & 26.6986 & 2669.86 & 32038.4 & 3.20384$\times 10^8$ \\
\hline
\multicolumn{6}{c}{}\\[-2mm]
\hline
\multicolumn{6}{c}{(b) Fixed $L=0.1$, varying $Q$}\\
\hline
$Q$ & $\omega_R$ &
$M_{\min}^{\rm LIGO}(M_\odot)$ &
$M_{\max}^{\rm LIGO}(M_\odot)$ &
$M_{\min}^{\rm LISA}(M_\odot)$ &
$M_{\max}^{\rm LISA}(M_\odot)$\\
\hline
 0.1 & 0.981333 & 26.3815 & 2638.15 & 31657.8 & 3.16578$\times 10^8$ \\
 0.3 & 0.99313 & 26.6986 & 2669.86 & 32038.4 & 3.20384$\times 10^8$ \\
 0.5 & 1.01925 & 27.4008 & 2740.08 & 32881. & 3.2881$\times 10^8$ \\
\hline
\end{tabular}
\end{table*}
Tables~\ref{tab:detectability0}--\ref{tab:detectability2} summarize the BH mass ranges for which the fundamental QNM lie within the sensitivity bands of current and future gravitational-wave detectors. The results are presented separately for  $s=0,\frac{1}{2},1,\frac{3}{2}$, and 2. For each field, panel (a) illustrates the effect of varying the  $L$ while keeping the charge fixed at $Q=0.3$, whereas panel (b) shows the dependence on the  charge $Q$ for a fixed value $L=0.1$.

It is evident that the detectable mass range is directly correlated with the real part of the quasinormal frequency. Since the oscillation frequency satisfies $\hat{f}\propto \omega_R/M$, larger values of $\omega_R$ correspond to larger black-hole masses producing gravitational-wave signals within the same detector band. Consequently, any parameter that increases the real part of the quasinormal frequency shifts the detectable mass interval toward higher masses, whereas a decrease in $\omega_R$ shifts it toward lower masses.

For the scalar  and electromagnetic  perturbations, increasing the  $L$ decreases the real part of the quasinormal frequency. As a result, both the minimum and maximum detectable masses for LIGO/Virgo and LISA decrease monotonically with increasing $L$. In contrast, for the Dirac, Rarita--Schwinger  and gravitational  perturbations, the real part of the quasinormal frequency increases with $L$, leading to a corresponding increase in the detectable BH mass range.

The influence of the electric charge is qualitatively similar for all perturbing fields considered in this work. As the charge $Q$ increases, the real part of the quasinormal frequency also increases, implying that progressively more massive BHs emit ringdown signals within the sensitivity windows of both ground-based and space-based detectors. Although the quantitative change is modest over the parameter range considered here, the trend is consistent across all spin fields.

Another noteworthy feature is the dependence on the spin of the perturbing field. The gravitational perturbations possess the largest oscillation frequencies and therefore correspond to the highest detectable BH masses, reaching several thousand solar masses for the LIGO/Virgo band and approximately $3\times10^8\,M_\odot$ for LISA. On the other hand, the Rarita--Schwinger ($s=\frac{3}{2}$) field exhibits the lowest oscillation frequencies among the fields studied, resulting in the smallest detectable mass ranges. The scalar, Dirac, and electromagnetic perturbations occupy intermediate regimes.

Overall, these results indicate that LSB in bumblebee gravity modifies the ringdown frequencies sufficiently to shift the astrophysical mass ranges accessible to gravitational-wave observations. While the changes are moderate for the values of $L$ considered here, future high-precision ringdown measurements from next-generation detectors may provide an independent means of constraining the LSB parameter through black-hole spectroscopy.
%
%
%
%

\section{Black Hole Thermodynamics}
Investigating BH thermodynamics is  important to understand how gravitation, quantum theory, and statistical physics are related. In this section, we investigate how modified gravity framework, such as bumblebee gravity affects the thermodynamic properties of the RN BH by analysing the influence of the Lorentz violating parameter $L$. The analysis of Hawking temperature of the BH allows us to examine how the modified space-time geometry effects the thermal behaviour of the BH \cite{ D. A. Gomes2020, I. Sakalli 2023, W. Liu2024, Gibbon77, ibungochouba2015}.
Hawking temperature is obtained from the surface gravity at the event horizon and serves as an important quantity for studying the evaporation process \cite{cardoso2001}. Also the entropy and heat capacity are also investigated to understand the thermodynamic stability of the RN-bumblebee black hole. The entropy provides information about the microscopic degrees of freedom associated with the event horizon, while the heat capacity governs the system's local thermodynamic stability. Specifically, stable configurations are characterized by a positive heat capacity, whereas negative values point to unstable states.
Therefore, above mentioned analysis will help us to explore the effects of LSB on the thermodynamic behaviour, phase transitions, and stability profiles of RN BHs within this modified gravitational context. 

\subsection{Hawking Temperature}\label{sec 5.1}
To evaluate the BH's thermodynamic characteristics, we begin by locating the event horizon, which corresponds to the largest real root of the metric function equation $A(r) = 0$
\begin{align}
r_{h_+}= M_+ \sqrt{M^2- \dfrac{2(1+L)}{2+L}Q^2}. 
\end{align}
Utilizing the surface gravity $\kappa_h$ at the event horizon, the corresponding Hawking temperature can be defined as \cite{I. M. Ablu2014, I. M. Ablu2010} 
 \begin{align} \label{temp}
 T_h = \dfrac{\kappa_h}{2\pi}.
 \end{align}
 We obtain the expression of Hawking temperature as
\textbf{ \begin{align}\label{temp2}
 T_h= \dfrac{(2+L)Mr_h- 2Q^2(1+L)}{2 \pi r_h^3\sqrt{1+L} (2+L) }.
 \end{align}}
 
\begin{figure}[h!]
\centering
  \subfloat[\centering ]{{\includegraphics[width=0.42\textwidth]{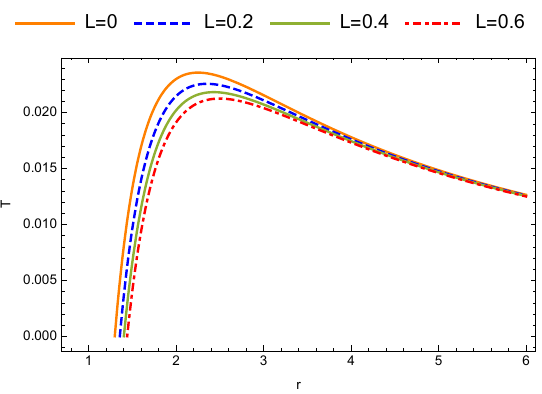}}\label{fig fpo}}
  \qquad
   \subfloat[\centering ]{{\includegraphics[width=0.42\textwidth]{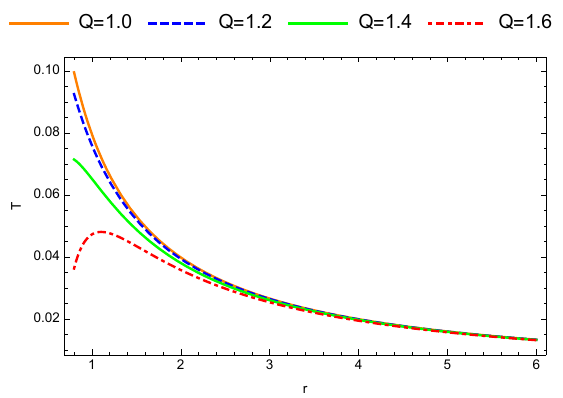}}\label{fig tempQ}}
   \caption{Plot of Hawking Temperature for different values of  (a) $L$,  and (b) $Q$ with horizon radius of BH.}
   \label{fig temp}
\end{figure}

 It is noted that, in the limit $L\rightarrow 0,$ we recover the standard RN result, providing a useful consistency check.
 In Fig. \ref{fig temp}, the variation of the Hawking temperature  with the horizon radius for the RN BHs in bumblebee gravity is plotted for different values of the  $L$  and $Q$   with fixed $M = 1$. From the plots, it is observed that increasing $L$ suppresses the maximum value of $T_h$, indicating that bumblebee gravity reduces the maximum temperature attained by the black hole.
Figure \ref{fig tempQ} displays the effect of varying the electric charge at fixed $L$, showing that higher values of $Q$ lead to a decline in the peak of Hawking temperature.

\subsection{Greybody Factors for  different Spin Fields}
Thermodynamic character of a BH is determined  physically through Hawking radiation, a quantum process by which the BH emits particles with a spectrum that is exactly thermal at the horizon. However, an observer situated at spatial infinity does not detect this pure blackbody spectrum. The radiation must traverse the curved space-time exterior to the horizon, which acts as an effective potential barrier for the propagating field modes. This barrier partially reflects and partially transmits the outgoing radiation. The fraction of radiation that escapes to infinity, is encoded in the GFs. The Hawking temperature obtained in the previous section characterizes the thermal nature of the radiation emitted at the event horizon. We now investigate the GFs for different spin fields. Since the effective potential depends explicitly on the spin of the emitted field, the transmission probability, and  the deviation from a perfect blackbody spectrum, are different for each spin fields.  The GF $\gamma_l^{(s)}$ quantifies 
the fraction of Hawking radiation of spin $s$ 
and frequency $\omega$ that is transmitted 
through the effective potential barrier 
$V_s(r)$ to reach a distant observer. 
Following the scattering problem setup, the 
wave function satisfies
\begin{align}
Z(r_*) &= T(\omega)\,e^{-i\omega r_*}, 
\qquad r_* \to -\infty, \\
Z(r_*) &= e^{-i\omega r_*} + 
R(\omega)\,e^{+i\omega r_*}, 
\qquad r_* \to +\infty.
\end{align}

Using the 6th order WKB approximation 
\cite{S.Iyer1987, konoplya2003}, the reflection 
coefficient is expressed as

\begin{equation}
R = \left(1 + e^{-2i\pi K}\right)^{-1/2},
\end{equation}
where $K$ satisfies
\begin{equation}
K - i\frac{\omega^2 - V_s(r_0)}
{\sqrt{-2V_s''(r_0)}} 
- \sum_{i=2}^{6}\Lambda_i(K) = 0,
\end{equation}

with $r_0$  marks the location where the effective potential reaches its maximum value, satisfying $V_s'(r_0) = 0$. The primes signify derivatives computed relative to the tortoise coordinate $r_*$, and $\Lambda_i$ stands for the WKB correction terms corresponding to the $i$-th order  \cite{konoplya2003}. The GF
is then defined as the transmission coefficient \cite{konoplya2019b,konoplya2019c,konoplya2020b}

\begin{align}
\gamma_l^{(s)}(\omega) = 
\left(1 + e^{2\pi i K_s}\right)^{-1},
\label{eq:GF}
\end{align}
where $K_s$ is determined from the effective 
potential $V_s(r)$ evaluated at its peak 
$r_0^{(s)}$ through the WKB expansion 
\cite{konoplya2003}. The same formula of
Eq.~(\ref{eq:GF}) is applied to each spin 
field $s = 0, \frac{1}{2}, 1, \frac{3}{2}, 2$ 
and the corresponding effective potential 
$V_s(r)$ is also derived. The resulting 
Hawking emission spectrum for all the spin is
\begin{align}
\frac{d^2E_s}{dt\,d\omega} = 
\frac{\gamma_l^{(s)}(\omega)}
{e^{\omega/T_h} - 1}
\cdot\frac{\omega}{2\pi}.
\label{eq:emission}
\end{align}
%
Equation (\ref{eq:emission}) 
establishes the direct connection between 
the perturbative dynamics through 
$\gamma_l^{(s)}$ and the thermodynamic 
sector through $T_h$.

\begin{figure*}[!t]
\centering

\subfloat[$s=0$]{%
\includegraphics[width=0.42\textwidth]{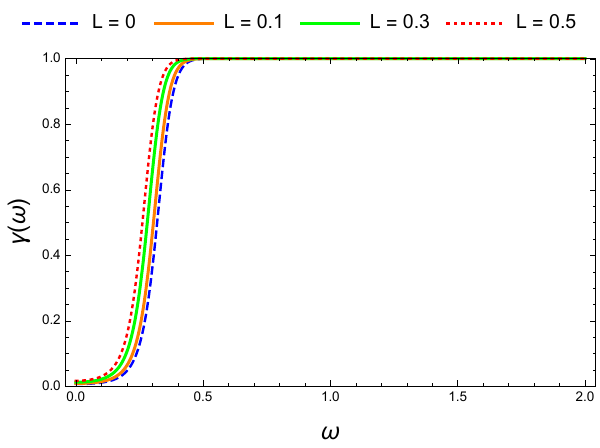}
\label{fig:GF1}}
\hfill
\subfloat[$s=\frac12$]{%
\includegraphics[width=0.42\textwidth]{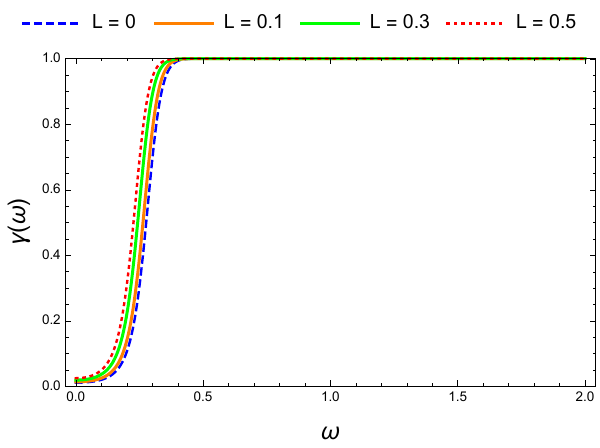}
\label{fig:GF2}}

\vspace{3mm}

\subfloat[$s=1$]{%
\includegraphics[width=0.42\textwidth]{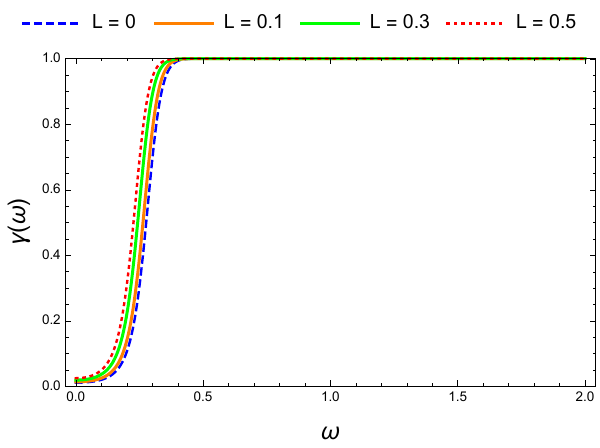}
\label{fig:GF3}}
\hfill
\subfloat[$s=\frac32$]{%
\includegraphics[width=0.42\textwidth]{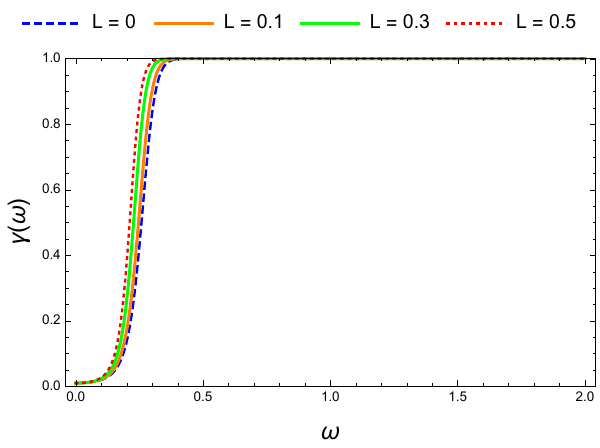}
\label{fig:GF4}}

\vspace{3mm}

\subfloat[$s=2$]{%
\includegraphics[width=0.42\textwidth]{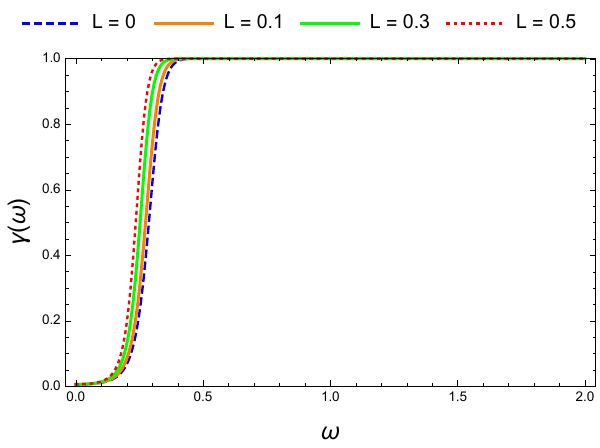}
\label{fig:GF5}}

\caption{Plot of the GF for different values of $L$ with (a) $s=0$, (b) $s=\frac12$, (c) $s=1$, (d) $s=\frac32$, and (e) $s=2$.}
\label{fig:GF_L}
\end{figure*}

 Figure \ref{fig:GF_L} presents the GF, $\gamma_{\ell}^{(s)}(\omega)$, as a function of frequency $\omega$ for fields of spin $s = 0, \frac{1}{2}, 1, \frac{3}{2}, 2$, evaluated for different values of  $L$. For all spin fields, the GF exhibits a qualitatively similar sigmoidal behavior, increasing monotonically from zero in the low-frequency regime to unity at high frequencies, indicating that higher-energy quanta more effectively overcome the effective potential barrier. An increase in $L$ systematically shifts the transmission onset toward higher frequencies for all spin fields. This behavior suggests that stronger Lorentz-violating effects enhance the effective potential barrier, thereby suppressing the transmission of Hawking quanta in the low-frequency regime. Similar qualitative trends are also observed across different spin fields, which implies that the dominant modification to the GF in the LSB background is primarily governed by the parameter $L$, rather than by the spin of the emitted field, and also highlights the universal impact of LSB on the underlying spacetime geometry. Since GF measures the transmission probability of Hawking radiation through the effective potential barrier, it naturally connects the dynamical scattering of perturbations with the thermodynamic emission properties of the BH.

\subsection{Heat Capacity}
To analyse the thermodynamic stability of the BH we compute the heat capacity \cite{kumar2020,priyo2023} for fixed $Q$ of the BH using the Hawking Temperature Eq. (\ref{temp2})  as
\begin{align} \label{eqn:CQ1}
C_Q= \left( \dfrac{dM}{dT_h}\right) _Q.
\end{align}
Expressing the mass of the BH in terms of $r_h$ and substituting it in Eq. (\ref{temp2}), we obtain 
\begin{align} \label{eqn:CQ2}
&M= \dfrac{r_h}{2}+\dfrac{(1+L)}{r_h(2+L)}Q^2  \hspace{0.2cm}  \rm{and} \nonumber\\ 
& T_h= \dfrac{(2+L)r_h^2-2(1+L)Q^2 }{4
(2+L) \sqrt{1+L}\pi r_h^3}.
\end{align}
The heat capacity of the BH can be calculated as
\begin{align}
C_Q=&\dfrac{dM/dr_h}{dT_h/dr_h} \nonumber \\
= &\dfrac{ 2\sqrt{1+L}(2(1+L)Q^2- (2+L)r_h^2)\pi r_h^4}{-6(1+L)Q^2r_h^2+(2+L)r_h^4}.
\end{align}

\begin{figure}[h!]
\centering
  \subfloat[\centering ]{{\includegraphics[width=0.42\textwidth]{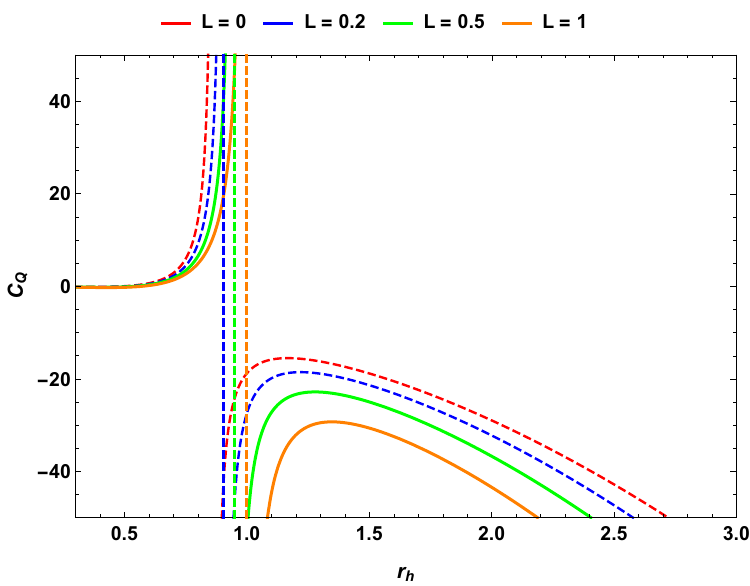}}\label{fig heatL}}
  \qquad
   \subfloat[\centering ]{{\includegraphics[width=0.42\textwidth]{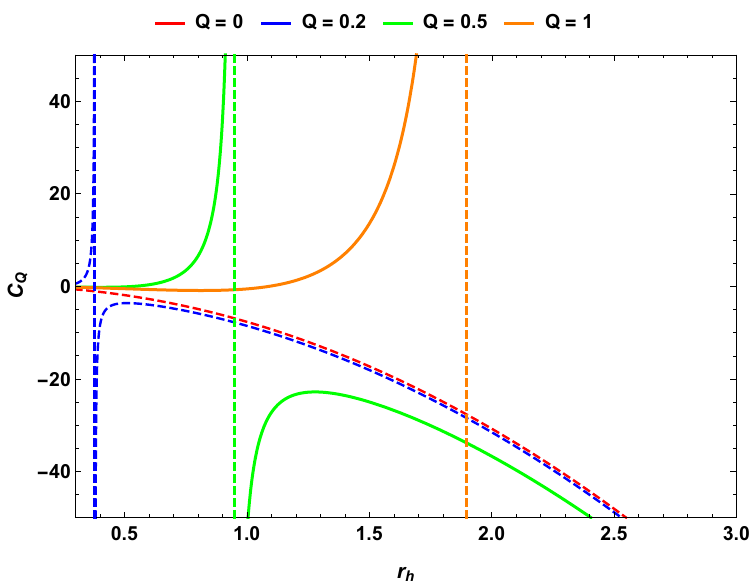}}\label{fig heatQ}}
   \caption{Plot of heat capacity $C_Q$ for different values of (a) $L$  and (b) $Q$}
   \label{fig heat}
\end{figure}
 
Figure \ref{fig heat} illustrates the behaviour of the heat capacity $C_Q$ 
as a function of $r_h$ for varying $L$ and $Q$. For all values of $L$, $C_Q$ diverges at the 
critical radius $r_h = r_c$, indicating a second-order phase 
transition. For $r_h < r_c$, the negative $C_Q$ indicates 
thermodynamic instability, where the BH spontaneously 
loses mass through Hawking evaporation, with increasing $L$ 
resulting instability. Beyond $r_c$, $C_Q$ becomes 
positive marking a stable thermodynamic phase, where 
increasing $L$ increases the stability of the BH. 
  Figure \ref{fig heatQ} describes the effect of charge $Q$. Increasing the value of $Q$ shifts the critical radius $r_c$ to larger values, indicating 
that higher charge demands a larger horizon size for 
thermodynamic stability. Comparing Figs. \ref{fig heatL} and  \ref{fig heatQ}, the effect 
of $Q$ on shifting $r_c$ has more impact  than that of $L$. We conclude that $Q$ plays a crucial role in controlling the phase structure of the RN BHs in 
bumblebee gravity.

\subsection{Entropy}
Using the 1st law of BH thermodynamics, the entropy of BH can be obtained,
\begin{align}
dS= \dfrac{dM}{T_h},
\end{align}
therefore,
\begin{align}
S= \int{\dfrac{dM}{T_h}},
\end{align}
where $S$ denotes the entropy of the BH
\begin{align}
S= \pi \sqrt{1+L}  r^2_h.
\end{align}
The corrected entropy of the black hole is given by
\begin{align}\label{eqn scorrected}
S_{corrected}=\pi \sqrt{1+L} r^2_h + \alpha \log  (\pi \sqrt{1+L}r_h^2).
\end{align}

\begin{figure}[h!]
\centering
  {{\includegraphics[width=0.42\textwidth]{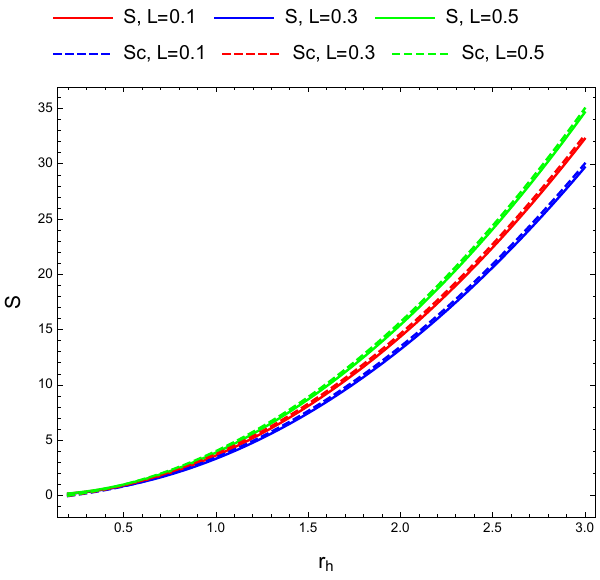}}\label{fig fpo}}
  \caption{Plot of the standard entropy $S$ and corrected entropy $S_c$ with horizon radius $r_h$ for different values of $L$.}
  \label{fig f12}
\end{figure}

Figure \ref{fig f12} illustrates the variation of entropy $S$ and corrected entropy $S_c$ as functions of the horizon radius $r_h$ for different values of the $L$. Both standard and corrected entropies exhibit monotonically increasing behaviour with $r_h$, consistent with the Bekenstein-Hawking area law. At small horizon, all curves nearly coincide, indicating the negligible effect of $L$. With increase in $r_h$, the curves separate for higher values of $L$ resulting larger entropy, showing the $L$ enhancing microscopic degrees of freedom of the BH. For all values of $L$, the corrected entropy $S_c$ (dashed curves) lies slightly below the standard entropy $S$ (solid curves), showing the effect of logarithmic thermal corrections which reduce the entropy. These results confirm that LSB enhances the BH entropy while logarithmic thermal corrections introduce only a small reduction, without altering the overall thermodynamic behaviour.

\subsection{Gibbs free energy}

Gibbs free energy combines the BH mass, the entropy-temperature term, and the electrostatic contribution from the charge. It characterizes the thermodynamic stability of the system at fixed temperature and charge, with its sign indicating whether the BH phase is favoured over thermal radiation. Gibbs free energy can be obtained from \cite{ding1,kastor1,ding2,jawad2017c}
\begin{align}\label{eqn gibbs1}
G= M- T_h S _{corrected}- \phi Q.
\end{align}
Substituting the corrected entropy $S_{corrected}$ obtained in Eq. \eqref{eqn scorrected}, along with the Hawking temperature $T_h$ and the electrostatic potential $\phi = Q/r_h,$ into Eq. \eqref{eqn gibbs1}, the explicit form of the Gibbs free energy for the LSB RN BH is obtained as
\begin{align}
G = & \dfrac{r_h}{2}+\dfrac{(1+L)}{r_h(2+L)}Q^2 - \dfrac{r_h^2-\frac{(1+L)}{r_h(2+L)}Q^2 }{4\pi r_h^3\sqrt{1+L}} \nonumber \\ 
&(\pi r^2_h \sqrt{1+L}  + \alpha \log  (\pi r_h^2) )- \dfrac{Q^2}{r_h}.
\end{align} 

\begin{figure}[h!]
\centering
  \subfloat[\centering ]{{\includegraphics[width=0.42\textwidth]{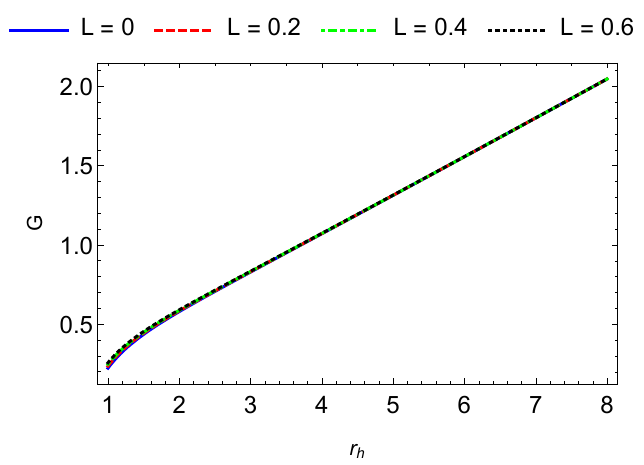}}\label{fig fpo}}
  \qquad
   \subfloat[\centering ]{{\includegraphics[width=0.42\textwidth]{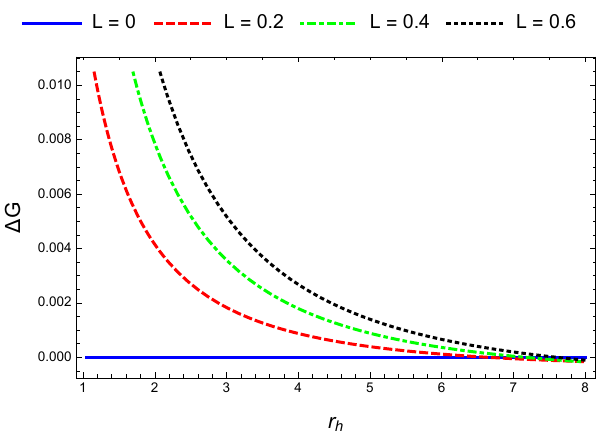}}\label{fig gibbs2}}
   \caption{Plot of Gibbs free energy $G$ with horizon radius $r_h$ for different values of the LSB parameter $L$ (a) $G$ vs $r_h$, (b) $\Delta G$ vs $r_h$.}
   \label{fig f13}
\end{figure}

As illustrated in Fig. \ref{fig f13}, the Gibbs free energy $G$ increases monotonically with increase in horizon radius $r_h$ for different values of the LSB parameter  $L$, indicating that  $L$ produces only a mild quantitative modification, as the curves closely overlapped. Since $G$ stays positive throughout, the pure thermal radiation phase is thermodynamically preferred over the BH state, which is consistent with the absence of a Hawking-Page transition in this non-AdS setup. Fig. \ref{fig gibbs2} plots $\Delta G = G(L)- G(0)$, isolating the correction due to LSB, which is largest at small $r_h$ and decreases toward zero as $r_h$ increases. This shows that LSB effects are significant for small, highly curved BHs and vanish for large BHs, where standard RN thermodynamics is recovered. Hence, the analysis shows LSB mainly affects the thermodynamic behaviour of small BHs, whereas the standard RN limit is smoothly recovered for large horizon radii.

\section{Discussion and Conclusion}

In this work, we have presented an integrated investigation into both the dynamical and thermodynamic behaviors of the RN BH within the framework of bumblebee gravity, where Lorentz symmetry is spontaneously broken. For the dynamical analysis, we successfully extended the unified Teukolsky master equation via the NP formalism to evaluate massless field perturbations of arbitrary spins ($s=0, 1/2, 1, 3/2, 2$). By casting the master equation into a one-dimensional Schr\"{o}dinger-like wave equation, we derived the effective potentials for each spin field. Our analysis revealed that the peak of the potential barrier decreases for higher-spin fields, indicating weaker confinement. Furthermore, we observed that the introduction of the LSB parameter $L$ weakens the potential barrier across all perturbing fields, whereas an increase in the BH charge $Q$ heightens the effective potential peak. We utilized both the Pad\'{e}-improved 6th-order WKB approximation and the AIM to compute the corresponding QNMs frequencies, providing a consistent numerical framework to study the damped oscillations of the modified spacetime.  The LSB parameter $L$ and the electric charge $Q$ produce distinct modifications to the quasinormal spectra. While increasing $Q$ generally enhances both the oscillation frequency and the damping rate for all perturbing fields, the effect of $L$ is spin dependent: it decreases the oscillation frequencies of scalar and electromagnetic perturbations but increases those of the Dirac, Rarita--Schwinger, and gravitational fields, whereas the damping rates decrease with increasing $L$ in all cases considered.

The observational implications of the computed QNMs were also examined by estimating the BH mass ranges detectable by LIGO/Virgo/KAGRA and LISA. We found that both the LSB parameter $L$ and the charge $Q$ produce systematic shifts in the oscillation frequencies and, consequently, in the detectable mass ranges. While the effect of $L$ depends on the spin of the perturbing field, increasing the charge generally enlarges the detectable mass range for all fields considered. These results suggest that future high-precision BH spectroscopy may offer an independent means of testing Lorentz-violating gravity through gravitational-wave observations.
Thermodynamically, the bumblebee gravity framework significantly alters the classical properties of the black hole. We established that increasing the LSB parameter $L$ or the electric charge $Q$ reduces the maximum Hawking temperature attained by the system. Evaluating the heat capacity $C_Q$ unveiled a second-order phase transition at a critical horizon radius $r_c$, marking the boundary between thermodynamic instability and stability. The charge $Q$ was found to play a dominant role in shifting this critical radius to larger values, directly controlling the phase structure of the modified RN BHs. Additionally, analyzing the entropy demonstrated that positive values of $L$ enhance the microscopic degrees of freedom, while the inclusion of logarithmic thermal corrections appropriately reduces the standard area-law entropy.

Ultimately, these combined results offer a comprehensive understanding of how Lorentz symmetry violation shapes the physical stability, wave propagation, and phase transitions of charged black holes. These findings provide a robust theoretical foundation and potentially observable signatures for testing modified gravity theories through future gravitational wave detections and black hole thermodynamic studies.

\onecolumn
\appendix
\section*{Appendix A}

This appendix summarizes the explicit forms of the AIM coefficients required for the numerical calculation of quasinormal modes. The coefficient $\lambda_0(x)$ is identical for all perturbation spins, whereas the coefficient $s^{n}_0(x)$ varies with the spin of the perturbing field due to the corresponding effective potential.

\begin{align}
\lambda_0(x)=& i \left(\dfrac{2 \sqrt{L+1} \omega
   }{\left(x-x_h\right) \kappa
   \left(x_h\right)}+\frac{2 \sqrt{L+1} \omega
   +i u'(x)}{u(x)}\right),\\
s^{0}_0(x)= &
\dfrac{1}{\sqrt{1+L}\,x^{3}(x-x_h)^2u(x)\kappa(x_h)^2} 
\left[ u(x)\left\lbrace(1+L)^{3/2}x^{3}\omega^{2}
-i(1+L)x^{3}\omega\,\kappa(x_h)
-\sqrt{1+L}\,(-1+2x)(x-x_h)^2\kappa(x_h)^2\right\rbrace  \right. \nonumber \\  & \left.
+x^{3}(x-x_h)\kappa(x_h)
\left\lbrace
\sqrt{1+L}(x-x_h)\lambda\,\kappa(x_h)
+(1+L)\omega\left(2\sqrt{1+L}\,\omega+i\,u'(x)\right)
\right\rbrace  \right],\\
s^{\frac{1}{2}}_0(x)= &\dfrac{\sqrt{L+1}} {4 x^2
   \left(x-x_h\right){}^2 u(x) \kappa
   \left(x_h\right){}^2} \left[x \left(x-x_h\right)
   \kappa \left(x_h\right)
   \left(\left(x-x_h\right) \kappa
   \left(x_h\right) \left((2 l+1)^2 \sqrt{L+1}
   x-2 u'(x)\right)  \right.\right.  \nonumber\\   & \left.\left.+4 x \omega  \left(2
   \sqrt{L+1} \omega +i u'(x)\right)\right)+4
   u(x) \left(-i x^2 \omega  \kappa
   \left(x_h\right)+\left(x-x_h\right){}^2
   \kappa \left(x_h\right){}^2+\sqrt{L+1} x^2
   \omega ^2\right)\right],\\
 s^{1}_0(x)= &  \frac{\left(x-x_h\right) \kappa
   \left(x_h\right) \left(\lambda  \sqrt{L+1}
   \left(x-x_h\right) \kappa
   \left(x_h\right)+(L+1) \omega  \left(2
   \sqrt{L+1} \omega +i
   u'(x)\right)\right)+(L+1) \omega  u(x)
   \left(\sqrt{L+1} \omega -i \kappa
   \left(x_h\right)\right)}{\sqrt{L+1}
   \left(x-x_h\right){}^2 u(x) \kappa
   \left(x_h\right){}^2},\\
s^{\frac{3}{2}}_0(x)= & \dfrac{\sqrt{L+1}}{4 x^2
   \left(x-x_h\right){}^2 u(x) \kappa
   \left(x_h\right){}^2}    \left[x \left(x-x_h\right)
   \kappa \left(x_h\right)
   \left(\left(x-x_h\right) \kappa
   \left(x_h\right) \left(\left(4 l^2+4
   l-3\right) \sqrt{L+1} x+2 u'(x)\right)   \right. \right. \nonumber\\ & \left. \left. +4 x
   \omega  \left(2 \sqrt{L+1} \omega +i
   u'(x)\right)\right)+4 u(x) \left(-i x^2
   \omega  \kappa
   \left(x_h\right)-\left(x-x_h\right){}^2
   \kappa \left(x_h\right){}^2+\sqrt{L+1} x^2
   \omega ^2\right)\right],\\
 s^{2}_0(x)= &  -\frac{\sqrt{L+1}}{(L+2) x^2
   \left(x-x_h\right){}^2 u(x) \kappa
   \left(x_h\right){}^2} \left[x \left(x-x_h\right)
   \kappa \left(x_h\right)
   \left(-\left(x-x_h\right) \kappa
   \left(x_h\right) \left(\sqrt{L+1} x
   \left(\lambda  (\lambda +2) (L+2)+6 (L+2)
   M    \right. \right.\right. \right. \nonumber\\ & \left. \left.           \left. \left.            -12 (L+1) Q^2 x\right)+6 (L+2)
   u'(x)\right)-(L+2) x \omega  \left(2
   \sqrt{L+1} \omega +i
   u'(x)\right)\right)-(L+2) u(x) \left(-i x^2
   \omega  \kappa \left(x_h\right)  \right. \right. \nonumber\\ & \left. \left.            -12
   \left(x-x_h\right){}^2 \kappa
   \left(x_h\right){}^2+\sqrt{L+1} x^2 \omega
   ^2\right)\right].
\end{align}

\twocolumn

\end{document}